\documentclass[aps,pra,twocolumn,groupedaddress,showpacs]{revtex4-1}

\usepackage{graphicx}
\usepackage{mathptmx}      
\usepackage{subfigure}
\usepackage{color}

\newif\ifSHOWCOMMENTS
\newif\ifACCEPTCOMMENTS

\SHOWCOMMENTSfalse     
\SHOWCOMMENTStrue

\ACCEPTCOMMENTStrue    
\ACCEPTCOMMENTSfalse   %

\DeclareRobustCommand\openone{\leavevmode\hbox{\small1\normalsize\kern-.33em1}}

\begin{document}


\title{Quantum Decoherence Scaling with Bath Size:\\
Importance of Dynamics, Connectivity, and Randomness}


\author{Fengping Jin}
\affiliation{Institute for Advanced Simulation, J\"ulich Supercomputing Centre,\\
Research Centre J\"ulich, D-52425 J\"ulich, Germany}
\author{Kristel Michielsen}
\affiliation{Institute for Advanced Simulation, J\"ulich Supercomputing Centre,\\
Research Centre J\"ulich, D-52425 J\"ulich, Germany}
\affiliation{%
RWTH Aachen University, D-52056 Aachen,
Germany
}%

\author{Mark Novotny}
\affiliation{Department of Physics and Astronomy, Mississippi State University,
Mississippi State, MS 39762-5167, USA}
\affiliation{HPC$^2$ Center for Computational Sciences, Mississippi State University,
Mississippi State, MS 39762-5167, USA}
\author{Seiji Miyashita}
\affiliation{
Department of Physics, Graduate School of Science,\\
University of Tokyo, Bunkyo-ku, Tokyo 113-0033, Japan
}
\affiliation{
CREST, JST, 4-1-8 Honcho Kawaguchi, Saitama, 332-0012, Japan}
\author{Shengjun Yuan}
\affiliation{Institute for Molecules and Materials, Radboud University of Nijmegen, \\
NL-6525AJ Nijmegen, The Netherlands}
\author{Hans De Raedt}
\affiliation{Department of Applied Physics, Zernike Institute for Advanced Materials,\\
University of Groningen, Nijenborgh 4, NL-9747AG Groningen, The Netherlands}


\date{\today}

\begin{abstract}
The decoherence of a quantum system $S$ coupled to a quantum
environment $E$ is considered.
For states chosen uniformly at random from the unit hypersphere in the Hilbert space
of the closed system $S+E$ we
derive a scaling relationship for the sum of the off-diagonal elements of
the reduced density matrix of $S$ as a function of the size $D_E$ of the
Hilbert space of $E$.
This sum decreases as $1/\sqrt{D_E}$ as long as
$D_E\gg 1$. This scaling prediction is tested by performing large-scale simulations which solve the
time-dependent Schr{\"o}dinger equation for a ring of spin-$1/2$ particles, four of them belonging to $S$ and the others to $E$.
Provided that the time evolution drives the whole system from the initial state
toward a state which has similar properties as states belonging to the class of quantum states
for which we derived the scaling relationship,
the scaling prediction holds. For systems which do not exhibit this feature, it is shown that increasing
the complexity (in terms of connections) of the environment or introducing a small amount of randomness in the interactions
in the environment suffices to observe the predicted scaling behavior.
\end{abstract}


\maketitle

\section{Introduction}

Decoherence of a quantum system $S$ interacting with a quantum
environment $E$ is of importance
for two reasons.
First,
decoherence of $S$ is the primary requirement for
$S$ to relax to a state described by a canonical ensemble
at a certain temperature~\cite{KUBO85}.
Second, decoherence is arguably the largest impediment for
practical, realizable quantum computers~\cite{NIEL00}.


The large interest in technological areas like spintronics, quantum computing and quantum information processing have stimulated the theoretical research of quantum dynamics in open and closed interacting systems. Besides this more application driven interest there persists the fundamental and still unanswered question under which conditions a finite quantum system reaches thermal equilibrium and how this can be derived from dynamical laws.

On the one hand there exists a variety of studies exploring the microcanonical thermalization in an isolated
quantum system~\cite{YUKA11,NEUM29,PERE84,DEUT91}. On the other hand there exist various studies
investigating the process of canonical thermalization of a system coupled to a (much) larger
system~\cite{TASA98,GOLD06,POPE06,REIM07,YUKA11,LIND09,SHOR11,REIM10} and of two finite identical
quantum systems prepared at different temperatures~\cite{PONO2011,PONO12}.

In previous work~\cite{YUAN09,JIN10b}, we numerically demonstrated that a
quantum system interacting with an environment at high temperature
relaxes to a state described by the canonical ensemble.
In this paper we focus on investigating the dynamic properties of the decoherence of
a quantum system $S$, being a subsystem of the whole system $S+E$.
We do this both with a theoretical prediction and
by simulating the dynamics of a relatively large system $S+E$ of spin-$1/2$ particles using a
time-dependent Schr{\" o}dinger equation (TDSE) solver~\cite{RAED06}.
In particular, we investigate the scaling of the degree of decoherence of $S$
with the size of $E$, keeping the size of $S$ fixed.
Based on similar arguments as given in Ref.~\cite{HAMS00}, we find that the degree of decoherence of
$S$ decreases as $1/\sqrt{D_E}$, where $D_E$ is the dimension of the
Hilbert space of the environment
if the state of the whole system is
chosen uniformly at random from the unit hypersphere in the Hilbert space.
In this paper, we denote states chosen uniformly at random from the unit hypersphere in the Hilbert space
of the whole system by ``$X$" and of the environment by ``$Y$".

We also address the question under what circumstances the whole system
evolves to a state which has the same degree of decoherence as a state ``$X$''.
In particular we study the case in which the initial
state of $S+E$ is a direct product of the state $\left| \uparrow\downarrow\uparrow\downarrow\right>$ of $S$ and a state ``$Y$'' of $E$.
If the initial state of the whole system $S+E$ is slightly different from a given state ``$X$'', the dynamics
may drive the whole system into a state which is very different from the given state ``$X$'', but which is of a similar type.
We investigate through our simulations when the
dynamics plays an important role in the decoherence in that
it can drive $S+E$ to a state ``$X$'' by introducing small world bond connections in $E$ and/or between $S$ and $E$
and by introducing randomness in the interaction strengths of the environment.

The paper is organized as follows.  In Section~II our theoretical
results for the scaling of the decoherence of $S$ are presented, together with
details of the one-dimensional ring of spin-$1/2$ particles which we simulate to
better understand the scaling prediction.  Sections~III-V contain results
for the one-dimensional rings under study. In particular we look at the effect of
adding additional bonds (Small World Bonds, SWBs) between the system and environment spins and/or between environment spins only
(Section~IV) and of
randomness in the interaction strengths of the Hamiltonian of the environment (Section~V).
Section~VI contains our conclusions and a
discussion of our results.

\section{Theory, Model, and Methods}
The time evolution of a closed quantum system is governed by the time-dependent
Schr{\" o}dinger equation (TDSE)~\cite{NEUM55,BALL03}.
If the initial density matrix of an isolated quantum system is non-diagonal then,
according to the time evolution dictated by the TDSE, it remains non-diagonal.
Therefore, in order to decohere the system $S$, it is necessary to have the
system $S$ interact with an
environment $E$, also called a heat bath or spin bath if the environment is composed of spins.
Thus, the Hamiltonian of the whole system $S+E$ takes the form
\begin{equation}
\label{Eq:H}
H=H_S + H_E + H_{SE}
\>,
\end{equation}
where $H_S$ and $H_E$ are the system and environment Hamiltonian respectively,
and $H_{SE}$ describes the interaction between the system and environment.
In what follows, we first describe the general theory that leads to the scaling of the decoherence of the system $S$ with the size
of $E$ and $S$.
We then describe in detail the spin-$1/2$ Hamiltonians we have simulated to provide a
case study for this scaling.

\subsection{Time evolution}
A pure state of the whole system $S+E$ evolves in time according to (in units of $\hbar=1$)
\begin{eqnarray}
\left|\Psi(t)\right\rangle
&=&
e^{-itH}\left|\Psi(0)\right\rangle=
\sum_{i=1}^{D_S} \sum_{p=1}^{D_E} c(i,p,t)\left|i,p\right\rangle
\>,
\label{eq4}
\end{eqnarray}%
where the set of states $\{ |i,p\rangle \}$ denotes
a complete set of orthonormal states in
some chosen basis, and $D_S$ and $D_E$ are the dimensions of the Hilbert spaces
of the system and
the environment, respectively.  We assume that $D_S$ and $D_E$ are both finite.

The spin Hamiltonian $H$ models a system with $N_S$ spin-$1/2$ particles and an environment with $N_E$ spin-$1/2$ particles.
Thus, $D_S=2^{N_S}$ and $D_E=2^{N_E}$.
The whole system $S+E$ contains $N=N_S+N_E$ spin-$1/2$ particles and the dimension of its Hilbert space is $D=D_SD_E$.
In our simulations we use the spin-up -- spin-down basis.
Numerically, the real-time propagation by $e^{-itH}$ is carried out by means
of the Chebyshev polynomial algorithm~\cite{TALE84,LEFO91,IITA97,DOBR03},
thereby solving the TDSE for the whole system starting from the initial state
$\left|\Psi(0)\right\rangle$.
This algorithm yields results that are very accurate
(close to machine precision), independent of the time step used~\cite{RAED06}.

\subsection{Computational aspects}
Computer memory and CPU time severely limit the sizes of the quantum systems
that can be simulated.
The required CPU time is mainly determined by the number of operations to be performed on
the spin-$1/2$ particles.
The CPU time does not put a hard limit on the simulation. However, the memory of the
computer does severely limit which system sizes can be calculated.
The state $\left|\Psi \right\rangle$ of a
$N$-spin-$1/2$ system is represented by a complex-valued vector of length $D = 2^N$.
In view of the potentially large number of arithmetic operations,
it is advisable to use 13 - 15 digit floating-point arithmetic
(corresponding to 8 bytes for a real number).
Thus, to represent a state of the quantum system of $N$ spin-$1/2$ particles on a conventional
digital computer, we need a least $2^{N+4}$ bytes.
Hence, the amount of memory that is required to simulate a quantum system with $N$ spin-$1/2$ particles
increases exponentially with $N$.
For example, for $N=24$ ($N=36$) we need at least 256~MB (1~TB) of memory to store a
single arbitrary state $\left|\Psi \right\rangle$.
In practice we need three vectors, memory for communication buffers,
local variables and the code itself.

The elementary operations performed by the computational kernel are of the form
$\left|\Psi \right\rangle \leftarrow U \left|\Psi \right\rangle$
where $U$ is a sparse unitary matrix with a very complicated structure
(relative to the computational basis).
Inherent to the problem at hand is that each operation $U$ affects all elements of the
state vector $\left|\Psi \right\rangle$ in a nontrivial manner.
This translates into a complicated scheme for accessing memory,
which in turn requires a sophisticated MPI communication scheme \cite{RAED07X}.

\subsection{Reduced density matrix}

The state of the quantum system $S$ is described by the reduced density matrix
\begin{equation}
\hat{\rho}(t)\equiv\mathbf{Tr}_{E}\rho \left( t\right)
\>,
\label{eq1}
\end{equation}
where $\rho \left( t\right) $ is the density matrix of the whole system $S+E$ at time $t$
and $\mathbf{Tr}_{E}$ denotes the trace over the degrees of freedom of the environment.
In terms of the expansion coefficients $c(i,p,t)$, the matrix element $(i,j)$
of the reduced density matrix reads
\begin{eqnarray}
\hat\rho_{ij}(t) &=&
\mathbf{Tr}_{E} \sum_{p=1}^{D_E}\sum_{q=1}^{D_E}
c^\ast(i,q,t)c(j,p,t)|j,p\rangle\langle i,q|
\nonumber \\
&=&\sum_{p=1}^{D_E} c^\ast(i,p,t)c(j,p,t)
\>.
\end{eqnarray}

We characterize the degree of decoherence of the system by
\begin{equation}
\sigma (t) =\sqrt{\sum_{i=1}^{D_S-1}\sum_{j=i+1}^{D_S}
\left\vert\widetilde\rho_{ij}(t) \right\vert ^{2}} \>,
\label{eqsigma}
\end{equation}%
where $\widetilde\rho_{ij}(t)$ is the matrix element $(i,j)$
of the reduced density matrix $\widetilde\rho$ in the representation
that diagonalizes $H_S$.
Clearly, $\sigma(t)$ is a global measure for the size of the
off-diagonal terms of the reduced density matrix in the representation
that diagonalizes $H_S$. If $\sigma(t)=0$ the system is in a state of full decoherence
(relative to the representation that diagonalizes $H_S$).

\subsection{Scaling property of $\sigma$}
We can prove a scaling property of $\sigma$ by assuming that the final state
of the whole system is a state ``$X$", a state that
is picked uniformly at random from the unit hypersphere in the Hilbert space.
The wave function of the whole system reads,
\begin{equation}
\left | \Psi \right > =
\sum_{i=1}^{D_S}\sum_{p=1}^{D_E} C_{i,p} \left | E_i^{(S)} \right >\left | E_p^{(E)} \right >,
\end{equation}%
where $\left \{ \left | E_i^{(S)} \right \rangle \right \}$
$\left(\left \{\left | E_p^{(E)} \right \rangle \right \}\right)$
is the set of eigenvectors of $H_S$ ($H_E$), and the real and imaginary parts of $C_{i,p}$
are real random variables.
The derivation of the scaling behavior follows Ref.~\cite{HAMS00}.
In particular Eqs.~(A8), (A12) and (A23) of Ref.~\cite{HAMS00} are used.
We introduce the following shorthand notation for the sum over the off-diagonal
elements, $\sum_{i\ne j}^{D_S}\kappa_{ij} =
\sum_{i=1}^{D_S}\sum_{j=1}^{D_S} \left(1-\delta_{ij}\right) \kappa_{ij}$
for any $\kappa_{ij}$, where $\delta_{ij}$ is the Kronecker delta function.
The expectation value is given by
\begin{widetext}
\begin{eqnarray}
\label{eq_scaling}
E\left(2\sigma^2\right)
&=&E\left(\sum_{i\neq j}^{D_S} \left |\sum_{p=1}^{D_E} C_{i,p}^* C_{j,p} \right|^2\right)
 = \sum_{i\neq j}^{D_S} \> \sum_{p=1,p'=1}^{D_E} E\left(
    C_{i,p}^* C_{j,p} C_{i,p'} C_{j,p'}^*\right) \cr
&=&\sum_{i\neq j}^{D_S} \> \sum_{p=1,p'=1}^{D_E} \left( \left(1-\delta_{p,p'}\right)
 E\left(C_{i,p}^* C_{j,p} C_{i,p'} C_{j,p'}^*\right)+\delta_{p,p'}
 E\left(C_{i,p}^* C_{j,p} C_{i,p'} C_{j,p'}^*\right) \right) \cr
&=&\sum_{i\neq j}^{D_S} \sum_{p=1}^{D_E} E\left(\left|C_{i,p}\right|^2
    \left|C_{j,p}\right|^2\right)
 = \sum_{i\neq j}^{D_S} \sum_{p=1}^{D_E} \frac{1}{D_S D_E\left(D_S D_E+1\right)} = \frac{D_S-1}{D_S D_E+1}=\frac{1-\frac{1}{D_S}}{D_E+\frac{1}{D_S}}
\> ,
\end{eqnarray}
\end{widetext}
where $E(\cdot)$ denotes the expectation value with respect to the probability
distribution of the
random variables $C_{i,p}$.
Equation~(\ref{eq_scaling}) does not require any condition on the
Hamiltonian Eq.~(\ref{Eq:H}).
For example, if $H_E$ is composed of two or more environments that do not
couple to each other, but only interact with the system, in
Eq.~(\ref{eq_scaling}) $D_E$ is the product of the sizes of the Hilbert spaces of all
the environments.  In addition,
Eq.~(\ref{eq_scaling}) does not impose any requirement on the
geometry.

From Eq.~(\ref{eq_scaling}) it follows that for any fixed value of $D_S>1$ and $D_E\gg 1$, $\sigma$ scales as
\begin{equation}
\label{Eq:ScaleFinal}
\sigma \approx
\frac{1}{\sqrt{2}}
\sqrt{E\left(2\sigma^2\right)}
=\frac{1}{\sqrt{2}}\sqrt{\frac{D_S-1}{D_SD_E+1}}
\sim \frac{1}{\sqrt{2D_E}}
\>.
\end{equation}
Therefore, if the size of the system $S$ is fixed (which is the case considered in this paper), $\sigma$ decreases as
$1/\sqrt{D_E}$ for large $D_E$.
Hence, for a spin-$1/2$ system $\sigma$ should decrease as
$2^{-N_E/2}$ for large $N_E$.

For fixed $D_S>1$, it follows from Eq.~(\ref{eq_scaling}) that the environment does not have to be very large for
Eq.~(\ref{Eq:ScaleFinal}) to hold,
which is in agreement with Ref.~\cite{GEMM06}.
Nevertheless, the existence of an environment is crucial.
If there is no environment, then the $\sigma$ approaches to a constant (see Appendix~\ref{AppA}), even if the whole system is initially in a state ``$X$''.

\begin{figure}[t]
\includegraphics[width=8cm]{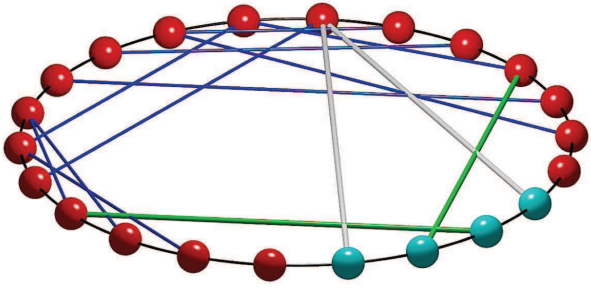}
\caption{\label{fig_lat}
(Color online) An example of a spin system used in the simulations.  The $N_S=4$ system
spin-$1/2$ particles are colored light gray (cyan), and the $N_E=18$ environment
spin-$1/2$ particles are colored dark gray (red).  The thin black segments show the connections for
a one-dimensional ring, which are the only bonds (interactions) present
in case~\uppercase\expandafter{\romannumeral1} and~\uppercase\expandafter{\romannumeral2} (see text).
The thick (green and white) bonds
show SWBs in $H_{SE}$. This particular example shows a spin system with $K=2$,
where $K$ denotes the maximum number of subsystem spins that are connected via SWBs with one environment spin
(thick white lines, see also Section~IV).  The medium thick (blue) bonds
show SWBs in $H_{E}$.}
\end{figure}

\subsection{Model and method}
For testing the predicted scaling of Eq.~(\ref{Eq:ScaleFinal}) we simulate
systems of spin-$1/2$ particles.
For studying the time evolution of the whole system $S+E$,
we consider a general quantum spin-$1/2$ model defined by the Hamiltonian
of Eq.~(\ref{Eq:H})
where
\begin{eqnarray}
\label{hamiltonian}
H_{S} &=&-\sum_{i=1}^{N_{S}-1}\sum_{j=i+1}^{N_{S}}\sum_{\alpha
=x.y,z}J_{i,j}^{\alpha }S_{i}^{\alpha }S_{j}^{\alpha }, \label{HAMS} \\ 
H_{E} &=&-\sum_{i=1}^{N_E-1}\sum_{j=i+1}^{N_E}\sum_{\alpha =x,y,z}\Omega
_{i,j}^{\alpha }I_{i}^{\alpha }I_{j}^{\alpha },  \label{HAME}\\ 
H_{SE} &=&-\sum_{i=1}^{N_{S}}\sum_{j=1}^{N_E}\sum_{\alpha =x,y,z}\Delta
_{i,j}^{\alpha }S_{i}^{\alpha }I_{j}^{\alpha }.  \label{HAMSE}
\end{eqnarray}%
Here, $S$ and $I$ denote the spin-$1/2$ operators
of the spins of the system and the environment, respectively (we use units such that
$\hbar$ and $k_B$ are one).
The spin components $S^\alpha_i$ and $I^\alpha_j$ are related to the Pauli spin matrices,
for example $S^x_i$ is a direct product of identity matrices and the Pauli
spin matrix
$\frac{1}{2}\sigma^x = \frac{1}{2}
\left(
\begin{array}{cc}
0 & 1 \cr
1 & 0
\end{array}
\right)$
in position $i$ of the direct product with
$1\le i\le N_S$.

For the geometry of the whole system, we focus on the one-dimensional ring
consisting of a system with $N_S=4$ spin-$1/2$ particles and an environment with $N_E$ spin-$1/2$ particles,
see Fig.~\ref{fig_lat}.
Past simulations have shown that a high connectivity spin-glass type of
environment is extremely efficient to decohere a system \cite{YUAN2006,YUAN09,YUAN11,BROX12},
so we may expect that the one-dimensional ring is one of the most difficult
geometries to obtain decoherence in short times.

We assume that the spin-spin interaction strengths of the system $S$ are isotropic,
$J_{i,j}^{\alpha }=J$ and that only the nearest-neighbor interaction strengths $\Omega_{i,j}^\alpha$
and $\Delta_{i,j}^\alpha$ are non-zero. Note that for a ring there are only two bonds with strength $\Delta_{i,j}^\alpha$ connecting $S$ and $E$.
We distinguish two cases:
\begin{itemize}
	\item Case~\uppercase\expandafter{\romannumeral1}: The non-zero values of $\Omega_{i,j}^\alpha$
and $\Delta_{i,j}^\alpha$ are generated uniformly at random from the range $[- \Omega, \Omega ]$
and $[- \Delta, \Delta ]$, respectively.
  \item Case~\uppercase\expandafter{\romannumeral2}: All non-zero values of the model parameters are identical, $\Omega_{i,j}^{\alpha }=J$ and
$\Delta_{i,j}^{\alpha }=J$. This corresponds to a uniform isotropic Heisenberg model
with interaction strength $J$.
\end{itemize}
We will see that these two cases show very different scaling properties of the
decoherence depending on the initial state.
We also investigate the effects of randomly adding small world bonds (SWBs)
between spins in the system and environment and between spins in the environment (see Fig.~\ref{fig_lat}).

The initial state of the whole system $S+E$ is prepared in two different ways,
namely:
\begin{itemize}
	\item ``$X$": We generate Gaussian random numbers $\{a(j,p), b(j,p)\}$ and set $c(j,p,t=0)=(a(j,p)+i b(j,p))/\sqrt{\sum_{j,p}(a^2(j,p)+b^2(j,p))}$.
	Clearly this procedure generates a point on the hypersphere in the $D$-dimensional Hilbert space.
	Alternatively, we generate points in the hypercube by using uniform random numbers in the interval $[-1,1]$.
	Our general conclusions do not depend on the procedure used (results not shown).
  \item $UDUDY$: The initial state of the whole system is a
product state of the system and environment. In this paper ($N_S=4$), we confine the discussion to the state $UDUDY$,
which means that the first, second, third, and fourth spin are in the up, down, up, and down state respectively,
and the state of the remaining spins is a ``$Y$" state in the $(D/2^4)$-dimensional Hilbert space.
The ``$Y$" state of the environment is prepared in the same way as the ``$X$" state of the whole system.
\end{itemize}

\section{Scaling analysis of $\sigma$}
All simulations are carried out for a system $S$ consisting of four spins
($N_S=4$) coupled to an
environment $E$ with the number of spins $N_E$ ranging from $2$ to $30$.
The interaction strengths $J_{i,i+1}^\alpha$ with
$1\le i\le N_S-1$ are always fixed to $J=-0.15$.
For case~\uppercase\expandafter{\romannumeral1} all non-zero
$\Omega_{i,j}^\alpha$ and $\Delta_{i,j}^\alpha$ are randomly generated
from the range $[-0.2,0.2]$.
For case~\uppercase\expandafter{\romannumeral2} all non-zero $\Omega_{i,j}^\alpha$ and
$\Delta_{i,j}^\alpha$ are equal to $J = -0.15$ (isotropic Heisenberg model).

\subsection{Verification of scaling: cases~\uppercase\expandafter{\romannumeral1} and~\uppercase\expandafter{\romannumeral2} with ``$X$"}
\label{sec3a}

\begin{figure}[t]
\includegraphics[width=8cm]{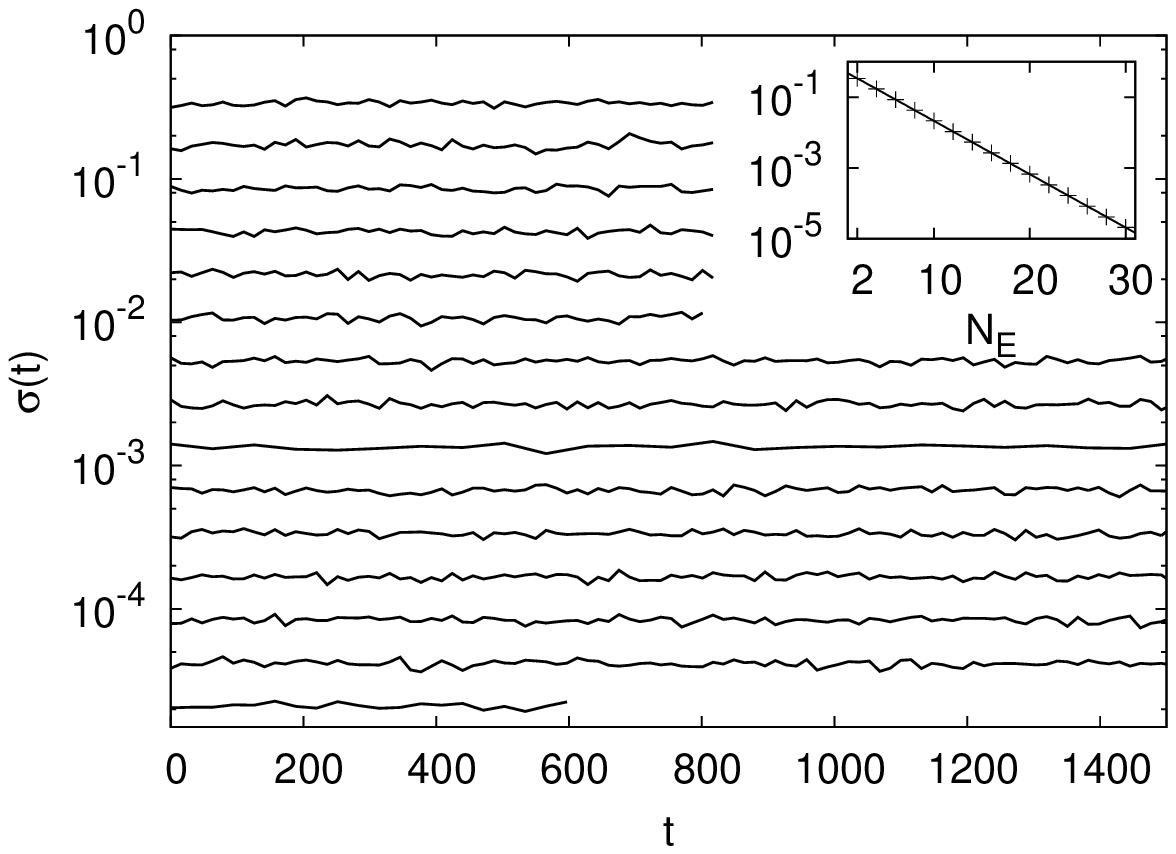}
\includegraphics[width=8cm]{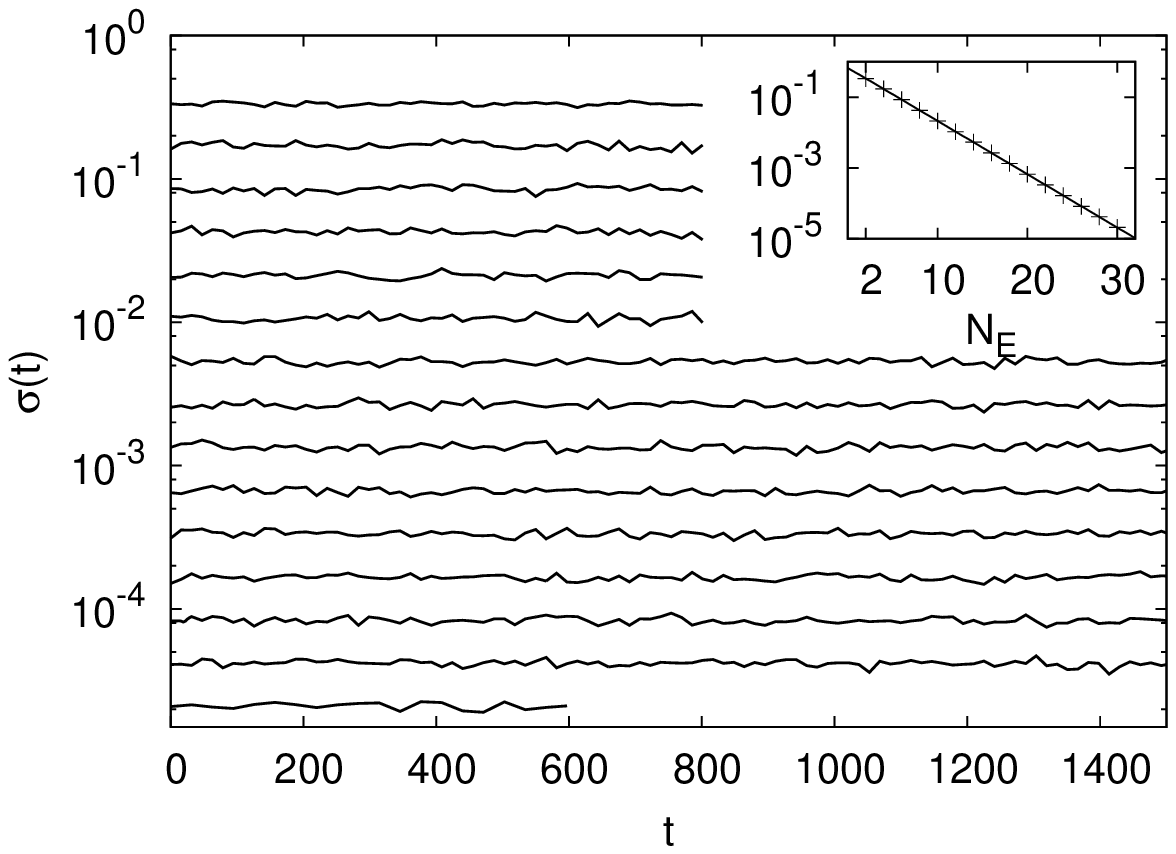}
\caption{\label{fig_case12}Simulation results for $\sigma(t)$
(see Eq.~(\ref{eqsigma})) for case~\uppercase\expandafter{\romannumeral1} (top) and
case~\uppercase\expandafter{\romannumeral2} (bottom) for different sizes $N=N_E+4$ of the whole system.
The initial state of the whole system is ``$X$" (see text).
Curves from top to bottom correspond to system sizes ranging from $N=6$ to $N=34$ in steps of $2$.
The insets show the time-averaged values of $\sigma(t)$ (pluses) as a function of the size $N_E$ of the environment,
confirming the theoretical prediction of Eq.~(\ref{Eq:ScaleFinal}) (solid line).
}
\end{figure}

We corroborate the scaling property of Eq.~(\ref{Eq:ScaleFinal}) by numerically
simulating the quantum spin system (see Eq.~(\ref{HAMS}) through (\ref{HAMSE})).
If we choose the initial state of the whole system to be an ``$X$" state, then during
the time evolution the whole system will remain in the state ``$X$".
Hence, the condition to derive Eq.~(\ref{Eq:ScaleFinal}) are fulfilled.
Fig.~\ref{fig_case12} demonstrates that the numerical results for both cases~\uppercase\expandafter{\romannumeral1} and~\uppercase\expandafter{\romannumeral2} agree
with Eq.~(\ref{Eq:ScaleFinal}).
In particular the insets in Fig.~\ref{fig_case12} show that for
both cases~\uppercase\expandafter{\romannumeral1} and~\uppercase\expandafter{\romannumeral2}, $\ln(2\sigma)\approx - N_E/2$,
and that $\sigma$ scales as $1/\sqrt{D_E}$ even if
$N_E=2$ and $N_S=4$ ($N_E<N_S$).

\subsection{Different initial conditions}
\label{sec3b}

We investigate the effects of the dynamics by preparing the initial state of the whole system such that it is
slightly different from ``$X$".
The initial state of the whole system is set to $UDUDY$.
In contrast to
Fig.~\ref{fig_case12}, we will see that the two cases~\uppercase\expandafter{\romannumeral1} and~\uppercase\expandafter{\romannumeral2} behave differently.

\subsubsection{Case~\uppercase\expandafter{\romannumeral1} and $UDUDY$}

\begin{figure}[t]
\includegraphics[width=8cm]{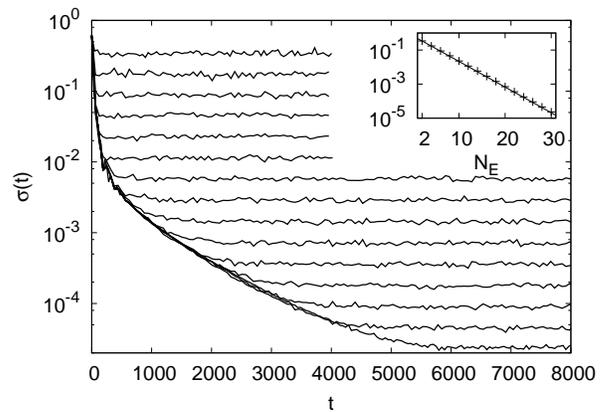}%
\caption{\label{fig_case1}
Simulation results for $\sigma(t)$ (see Eq.~(\ref{eqsigma})) for case~\uppercase\expandafter{\romannumeral1} for different sizes $N=N_E+4$ of the whole system.
The initial state of the whole system is $UDUDY$ (see text).
Curves from top to bottom correspond to system sizes ranging from $N=6$ to $N=34$ in steps of $2$.
The inset shows the time-averaged values of $\sigma(t)$ (pluses) as a function of the size $N_E$
of the environment. The data obey the scaling property of Eq.~(\ref{Eq:ScaleFinal}) (solid line).
}
\end{figure}

In Fig.~\ref{fig_case1}, we present the simulation results for case~\uppercase\expandafter{\romannumeral1}, the
couplings in the Hamiltonians $H_E$ and $H_{SE}$ are chosen uniformly at random.
The size $N=N_E+4$ of the whole system ranges from $6$ to $34$.
An average over the long-time stationary steady-state values of $\sigma(t)$ still obeys the
scaling property of
Eq.~(\ref{Eq:ScaleFinal}), showing that $\sigma$ decreases as $1/\sqrt{D_E}$, where $D_E=2^{N_E}$.
If $N_E\rightarrow\infty$, $\sigma\rightarrow0$. This suggests that in the thermodynamical limit the system $S$
decoheres completely.

\subsubsection{Case~\uppercase\expandafter{\romannumeral2} and $UDUDY$}

\begin{figure}[t]
\includegraphics[width=8cm]{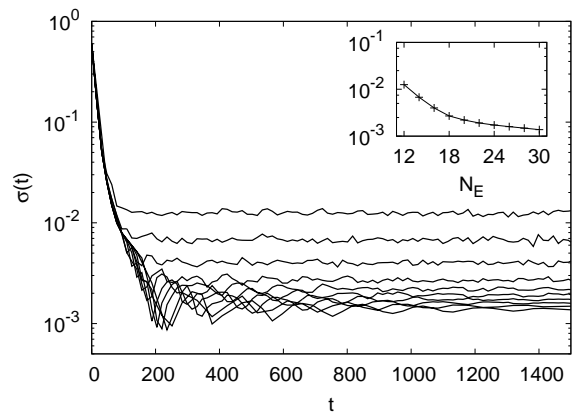}%
\caption{\label{fig_case2}Same as Fig.~\ref{fig_case1} for case~\uppercase\expandafter{\romannumeral2} instead of case~\uppercase\expandafter{\romannumeral1}.
Curves from top to bottom correspond to system sizes ranging from $N=16$ to $N=34$ in steps of $2$.
The solid line in the inset is a guide to the eyes.
}
\end{figure}

We consider the case in which the whole system is described by the isotropic Heisenberg
model ($J^\alpha_{i,i+1}=\Omega^\alpha_{i,i+1}=\Delta^\alpha_{i,i+1}=J$).
In Fig.~\ref{fig_case2} we present simulation results
for different system sizes $N=N_E+4$ ranging from $16$ to $34$.
From Fig.~\ref{fig_case2}, it is seen that the behavior for case~\uppercase\expandafter{\romannumeral2}
is totally different from that of case~\uppercase\expandafter{\romannumeral1} (see Fig.~\ref{fig_case1}).
In particular, $\sigma(t)$ does not scale with the dimension of the environment.
From the present numerical results, we cannot make any conclusions about the limit
for large $N_E$. However if $\sigma(t)$ approaches zero as $N_E\rightarrow\infty$
(see the fifth column of Table~\ref{tab1}) it
does so very slowly.

\subsection{Computational effort}
In this paper, the largest number of spins that we simulated is $N=34$.
Using the Chebyshev polynomial algorithm and a large time step ($\tau\approx 10\pi$),
the $N=34$ simulation for the bottom curves in
Fig.~\ref{fig_case12} (up to a time $t \approx 600$) took about $0.3$ million core hours on
$16384$~BG/P (IBM Blue Gene P) processors, using $1024$~GB of memory.
Similarly, it took about $4$ million core hours to complete the $N=34$ curve in Fig.~\ref{fig_case1} (up to a time $t\approx 8000$).

\subsection{Summary: initial state dependence}

\begin{figure}[t]
\includegraphics[width=8cm]{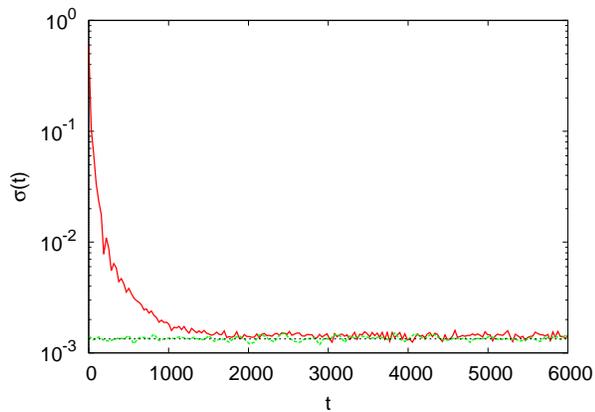}%
\caption{\label{fig_22spin}
(Color online) Simulation results for $\sigma(t)$ for case~\uppercase\expandafter{\romannumeral1} for $N=22$ and $N_S=4$.
Red solid line: the initial state is $UDUDY$ (see text); green dashed line: the initial state is ``$X$" (see text).
}
\end{figure}
\begin{figure}[t]
\includegraphics[width=8cm]{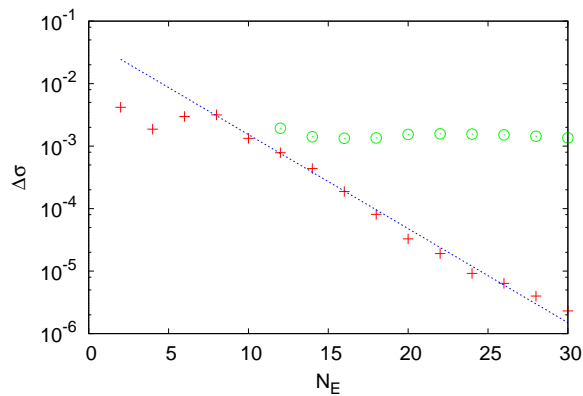}
\caption{\label{fig_diffsigma}
(Color online) Difference $\Delta\sigma$ between the time-averaged values of $\sigma(t)$
for the initial state $UDUDY$ and ``$X$" of the whole system (see Table~\ref{tab1})
as a function of the size of the environment $N_E$.
Pluses: case~\uppercase\expandafter{\romannumeral1}; circles: case~\uppercase\expandafter{\romannumeral2}.
The dotted line is a linear fit to the data (pluses) for the $UDUDY$ initial state,
excluding the first three data points, resulting in $\Delta\sigma=0.049/\sqrt{D_E}$.
}
\end{figure}

\squeezetable
\begin{table}
\caption{\label{tab1}
The time average of $\sigma(t)$ in the stationary regime shown in
Figs.~\ref{fig_case12}, \ref{fig_case1} and
\ref{fig_case2}.
}
\begin{ruledtabular}
\begin{tabular}{c|c|cc|cc}
 & prediction    & \multicolumn{2}{c|}{case~\uppercase\expandafter{\romannumeral1}} & \multicolumn{2}{c}{case~\uppercase\expandafter{\romannumeral2}} \\
$N_E$& of Eq.~(\ref{Eq:ScaleFinal}) & $UDUDY$ & ``$X$" & $UDUDY$ & ``$X$"\\
\hline\noalign{\smallskip}
2 & $3.397\times 10^{-1}$ & $3.416\times 10^{-1}$ &	$3.375\times 10^{-1}$ &            					    & $3.334\times 10^{-1}$ \\
4 & $1.708\times 10^{-1}$ & $1.746\times 10^{-1}$	& $1.727\times 10^{-1}$ &						              & $1.711\times 10^{-1}$ \\
6 & $8.554\times 10^{-2}$ & $8.834\times 10^{-2}$	& $8.536\times 10^{-2}$ &                         & $8.492\times 10^{-2}$ \\
8 & $4.279\times 10^{-2}$ & $4.598\times 10^{-2}$	& $4.282\times 10^{-2}$ &                         & $4.265\times 10^{-2}$ \\
10& $2.139\times 10^{-2}$ & $2.286\times 10^{-2}$	& $2.153\times 10^{-2}$ &                         & $2.121\times 10^{-2}$ \\
12& $1.070\times 10^{-2}$	& $1.149\times 10^{-2}$ & $1.071\times 10^{-2}$ &	$1.254\times 10^{-2}$		& $1.061\times 10^{-2}$	\\
14&	$5.349\times 10^{-3}$ & $5.795\times 10^{-3}$ & $5.357\times 10^{-3}$ &	$6.756\times 10^{-3}$		& $5.346\times 10^{-3}$ \\
16&	$2.674\times 10^{-3}$ & $2.866\times 10^{-3}$	& $2.678\times 10^{-3}$ & $3.997\times 10^{-3}$	  & $2.663\times 10^{-3}$ \\
18&	$1.337\times 10^{-3}$ & $1.430\times 10^{-3}$ & $1.349\times 10^{-3}$ &	$2.694\times 10^{-3}$ 	& $1.343\times 10^{-3}$ \\
20&	$6.686\times 10^{-4}$ & $7.065\times 10^{-4}$ & $6.736\times 10^{-4}$ &	$2.204\times 10^{-3}$		& $6.641\times 10^{-4}$ \\
22&	$3.343\times 10^{-4}$ & $3.542\times 10^{-4}$ & $3.352\times 10^{-4}$ &	$1.909\times 10^{-3}$		& $3.347\times 10^{-4}$ \\
24&	$1.672\times 10^{-4}$ & $1.766\times 10^{-4}$ & $1.674\times 10^{-4}$ &	$1.722\times 10^{-3}$		& $1.658\times 10^{-4}$ \\
26&	$8.358\times 10^{-5}$ & $9.005\times 10^{-5}$ & $8.368\times 10^{-5}$ &	$1.599\times 10^{-3}$		&	$8.283\times 10^{-5}$	\\
28&	$4.179\times 10^{-5}$ & $4.551\times 10^{-5}$ & $4.151\times 10^{-5}$ &	$1.481\times 10^{-3}$		& $4.176\times 10^{-5}$ \\
30& $2.089\times 10^{-5}$ & $2.338\times 10^{-5}$ & $2.107\times 10^{-5}$ & $1.379\times 10^{-3}$		& $2.104\times 10^{-5}$ \\
\end{tabular}
\end{ruledtabular}
\end{table}

For an initial state ``$X$" of the whole system the scaling of $\sigma$,
as given by Eq.~(\ref{Eq:ScaleFinal}), works extremely well
for both case~\uppercase\expandafter{\romannumeral1} and case~\uppercase\expandafter{\romannumeral2},
as seen in Fig.~\ref{fig_case12}.
When the initial state is $UDUDY$, we can understand the very different behavior
of cases ~\uppercase\expandafter{\romannumeral1} and~\uppercase\expandafter{\romannumeral2}, see Figs.~\ref{fig_case1} and~\ref{fig_case2}, by considering the stationary states that are obtained.
Figure~\ref{fig_22spin} shows that the final values of $\sigma(t)$ for case~\uppercase\expandafter{\romannumeral1}
are very close for both initial states ``$X$" and $UDUDY$.
This suggests that the final stationary state in case~\uppercase\expandafter{\romannumeral1} has properties similar to those of a state ``$X$",
and hence case~\uppercase\expandafter{\romannumeral1} obeys the scaling property
of Eq.~(\ref{Eq:ScaleFinal}) to a good approximation.
The time-averaged values of $\sigma(t)$ in Figs.~\ref{fig_case12},
\ref{fig_case1} and \ref{fig_case2}, denoted by $\overline{\sigma}$, are listed in Table~\ref{tab1}.
From Table~\ref{tab1}, we see that the values of $\overline{\sigma}$ for case~\uppercase\expandafter{\romannumeral2} with an
initial state $UDUDY$ are very different from those with an initial state ``$X$",
and do not show the scaling property of Eq.~(\ref{Eq:ScaleFinal}).
Thus, the numerical results suggest that the initial state and the randomness of the interaction strengths
play a very important role in
the dynamical evolution of the decoherence of a system coupled to an environment.
In particular, for case~\uppercase\expandafter{\romannumeral2}, starting from a state ``$X$"
the time-averaged values of $\sigma(t)$ scale as
$\overline{\sigma} \approx 1/\sqrt{D_E}$, but such scaling is not observed for starting from a state $UDUDY$.

From Table~\ref{tab1}, it is seen that the values of $\overline{\sigma}$ for case~\uppercase\expandafter{\romannumeral1} with the initial state $UDUDY$ are always
slightly larger than those with the initial state ``$X$".
Therefore, it is interesting to examine the difference $\Delta\sigma$ between the values of
$\overline{\sigma}$ for the initial states $UDUDY$ and ``$X$".
Figure~\ref{fig_diffsigma} shows that $\Delta\sigma$ for case~\uppercase\expandafter{\romannumeral1} (red pluses)
also scales as $1/\sqrt{D_E}$ (dotted line),
except for the first three data points,
which is probably due to large fluctuations in the calculations for these small system sizes.
Therefore, the dynamics of case~\uppercase\expandafter{\romannumeral1} will drive the system to a state ``$X$"
only when the environment approaches infinity.
Figure~\ref{fig_diffsigma} also shows that $\Delta\sigma$ for case~\uppercase\expandafter{\romannumeral2} (circles)
is almost constant for system sizes $N$ ranging from $16$ to $34$.
Hence, it is unlikely that case~\uppercase\expandafter{\romannumeral2} with the initial state $UDUDY$ will decohere, even if the
simulations could be performed for much longer times and for larger system sizes.

\section{Connectivity: ring with small world bonds}
\label{sec3c}

\begin{figure*}[t]
\includegraphics[width=8cm]{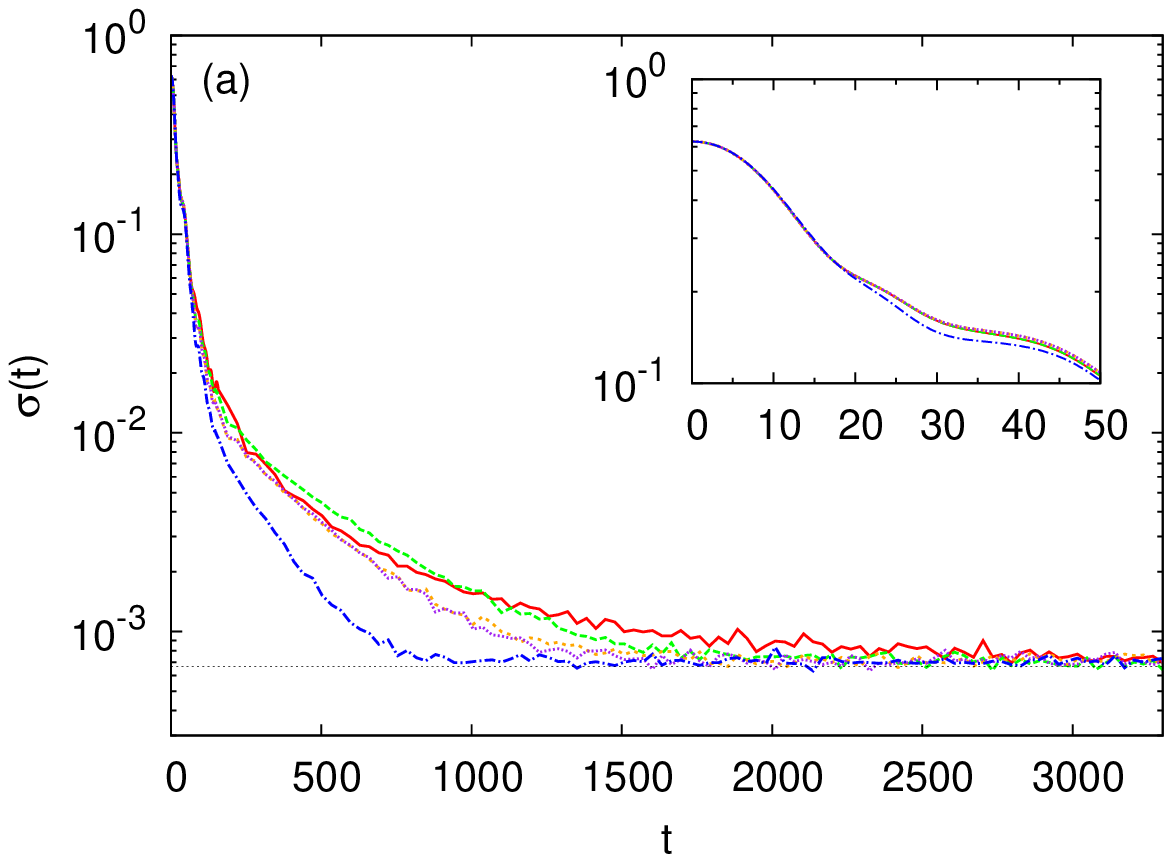}\includegraphics[width=8cm]{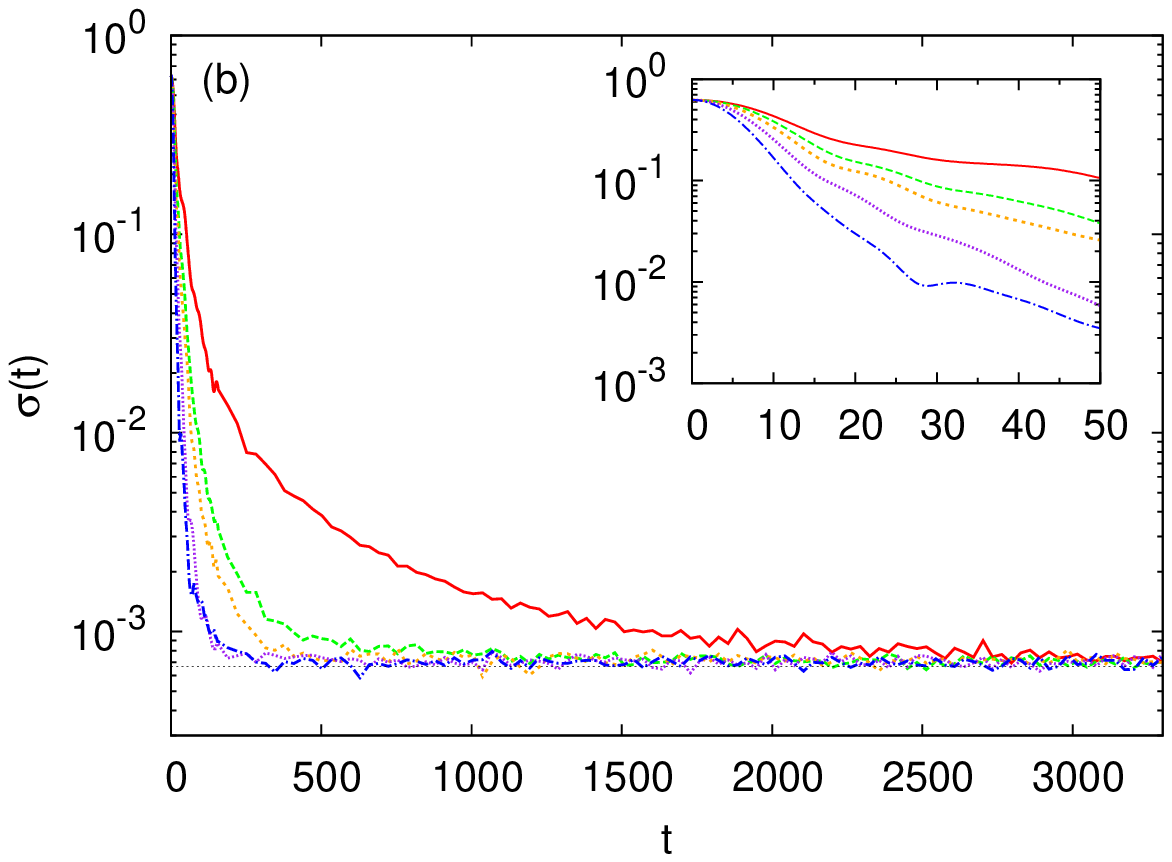}
\includegraphics[width=8cm]{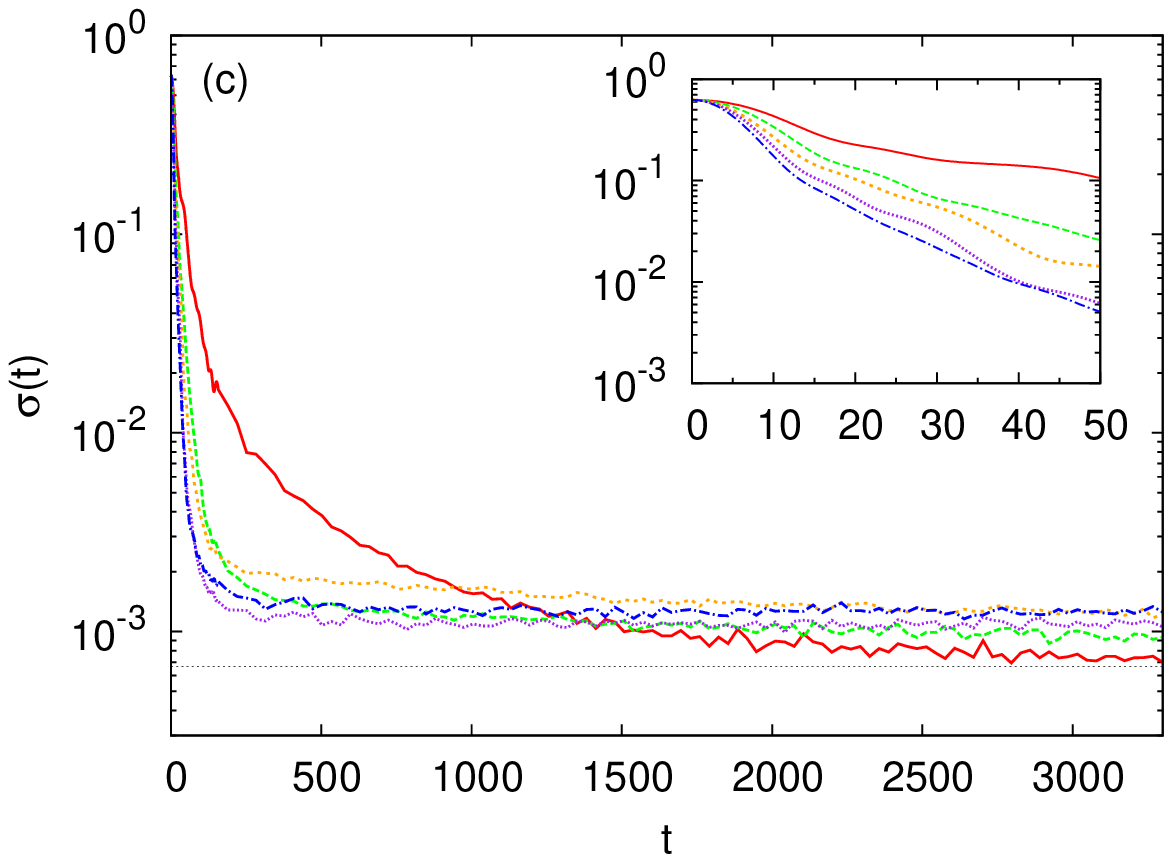}\includegraphics[width=8cm]{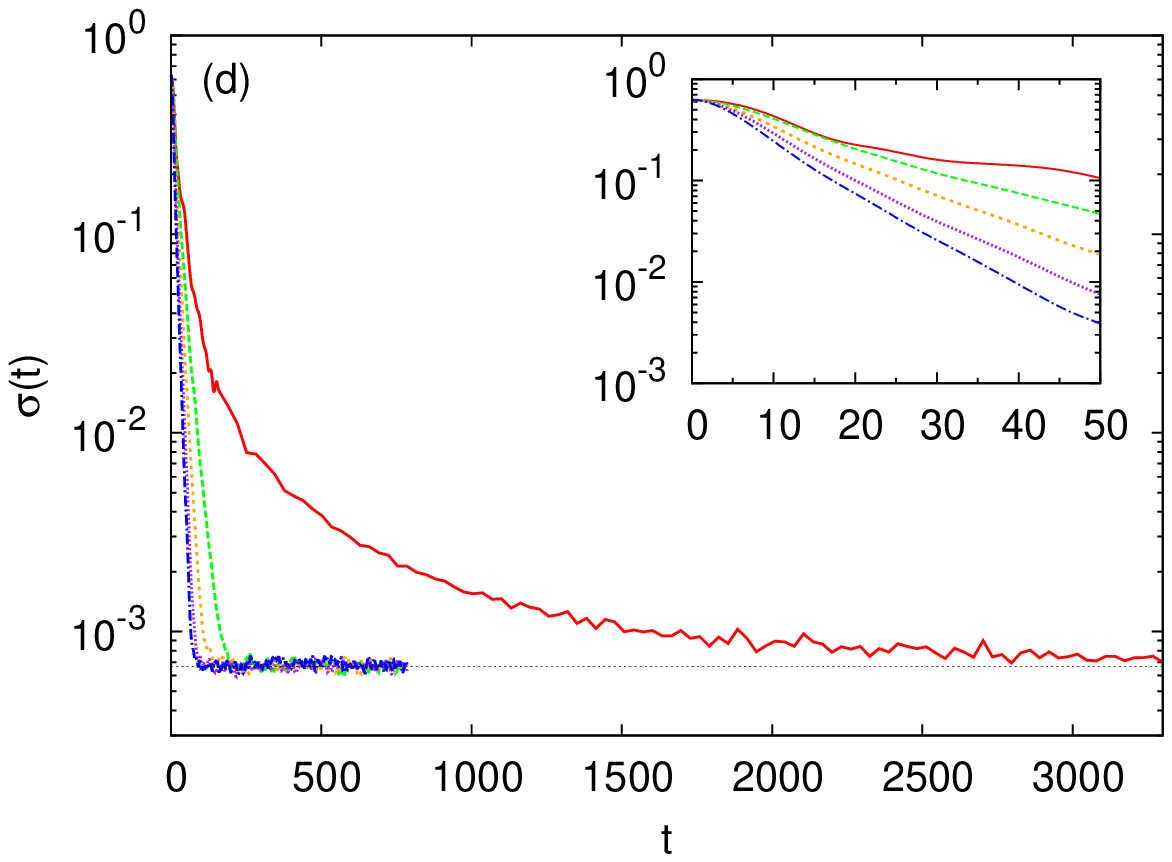}
\caption{\label{fig_case1_swb}
(Color online) Simulation results of $\sigma(t)$ for case~\uppercase\expandafter{\romannumeral1} with
$N=24$ and $N_S=4$ with SWBs added.
The initial state is $UDUDY$.
The dotted horizontal line represents the value of Eq.~(\ref{Eq:ScaleFinal}).
Red solid line: ring without SWBs.
(a) Randomly added SWBs between non-neighboring environment spins.
Green long-dashed line: one SWB;
orange dotted line: two SWBs;
purple short-dashed line: four SWBs;
blue dotted-dashed line: eight SWBs.
(b) Randomly added SWBs between the system and environment spins such that $K=1$.
Green long-dashed line: one SWB;
orange dotted line: two SWBs;
purple short-dashed line: four SWBs;
blue dotted-dashed line: eight SWBs.
(c) Randomly added SWBs between the system and environment spins such that $K=2$.
Green long-dashed line: two SWBs;
orange dotted line: four SWBs;
purple short-dashed line: six SWBs;
blue dotted-dashed line: eight SWBs.
(d) Same as (c) except that each pair of non-neighboring environment spins is connected by a SWB.
Insets: time evolution for short times.
}
\end{figure*}

\begin{figure*}[t]
\includegraphics[width=8cm]{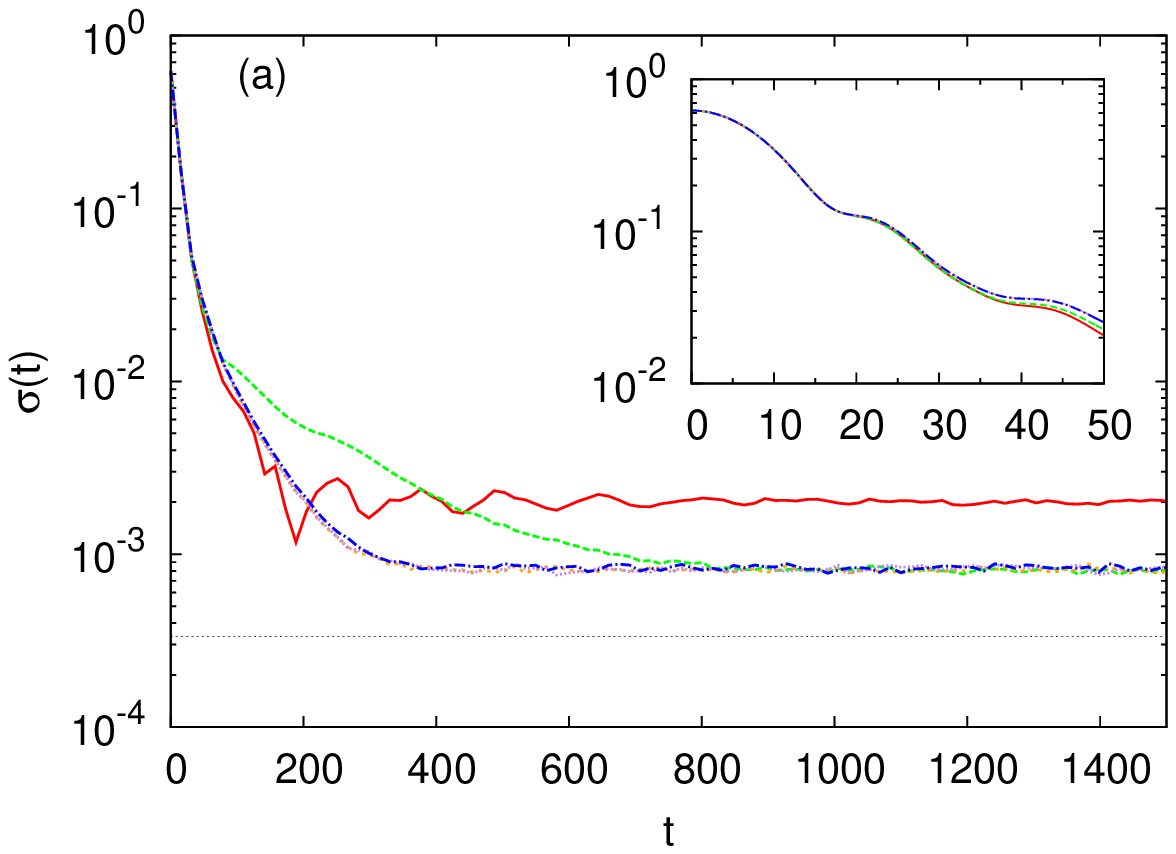}\includegraphics[width=8cm]{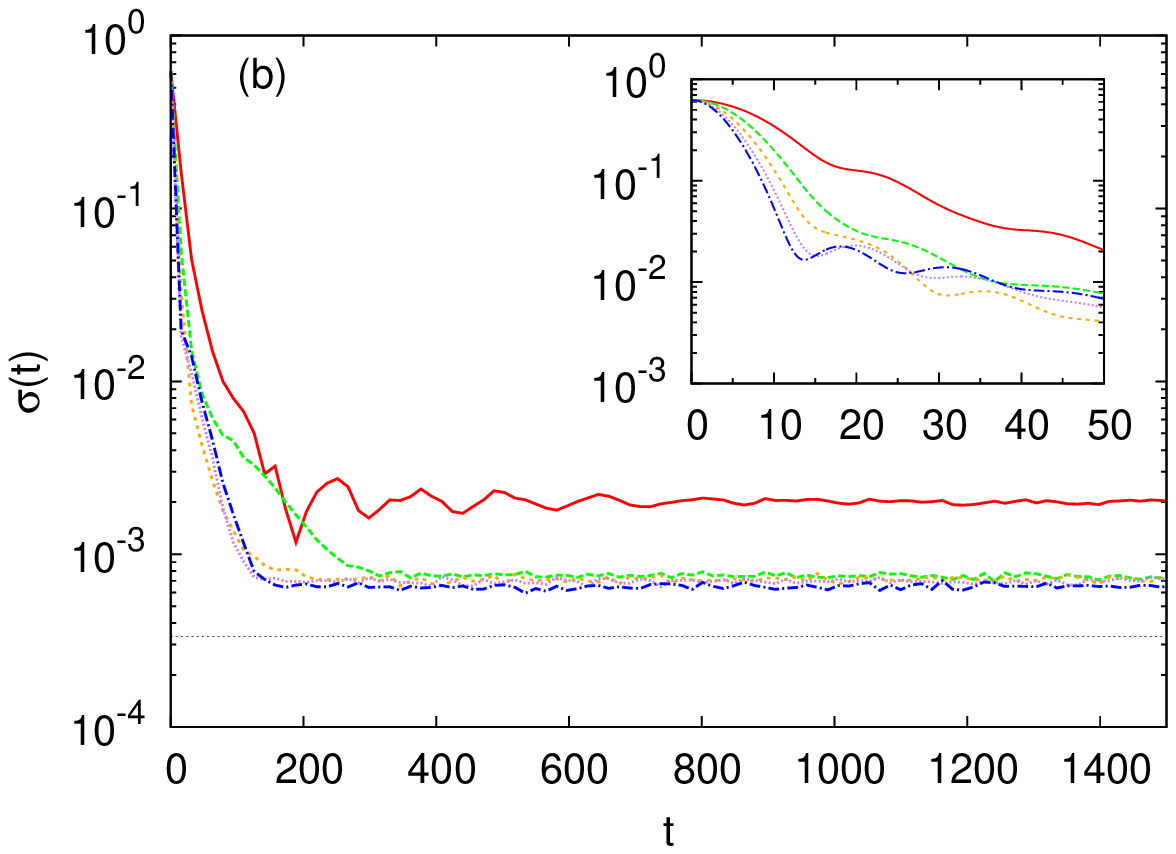}
\includegraphics[width=8cm]{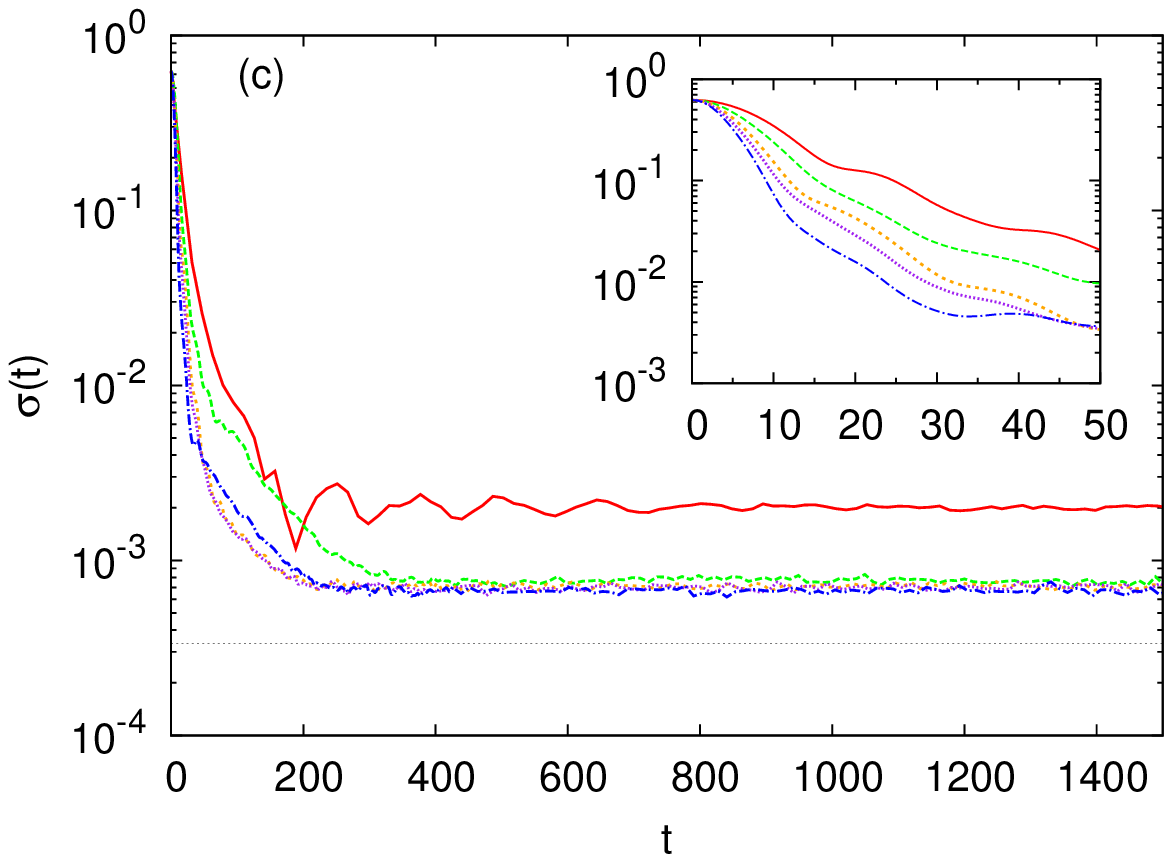}\includegraphics[width=8cm]{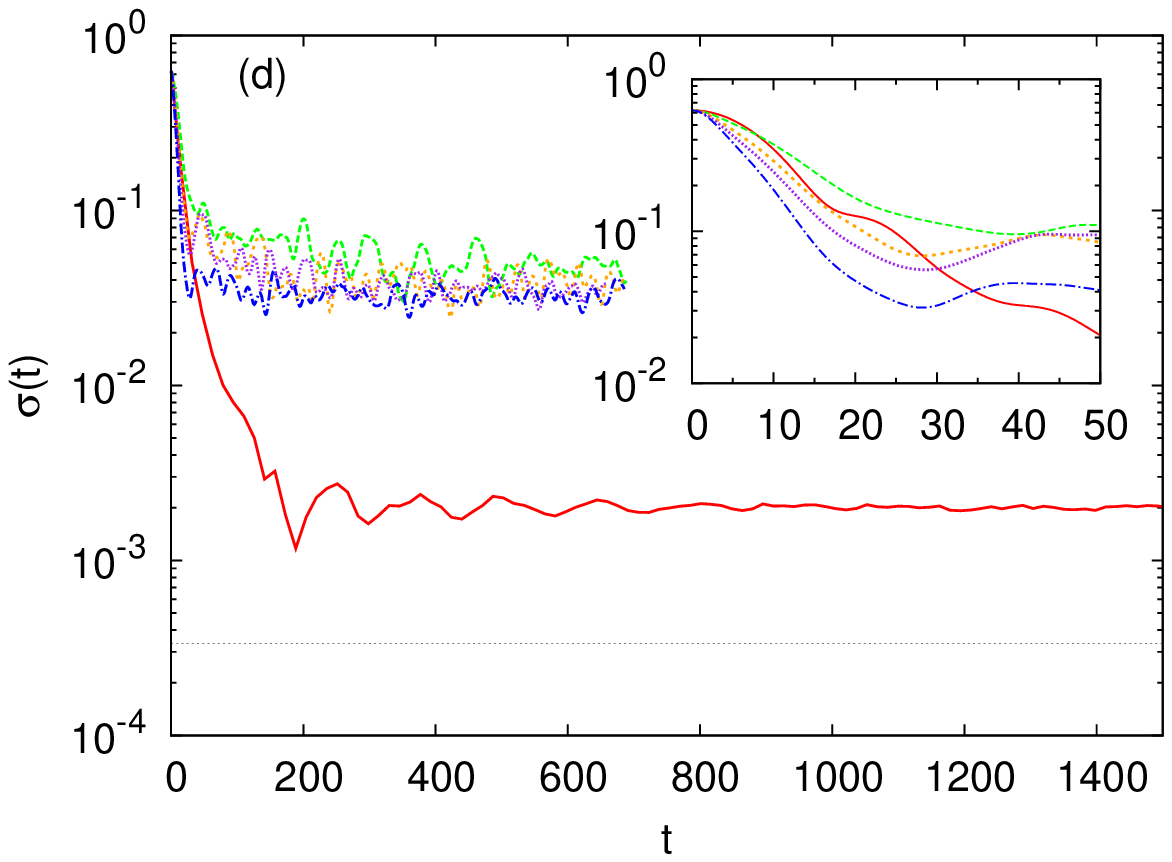}
\caption{\label{fig_case2_swb}
(Color online) Simulation results of $\sigma(t)$ for case~\uppercase\expandafter{\romannumeral2} with $N=26$ and $N_S=4$
with isotropic SWBs added.
The initial state is $UDUDY$.
The dotted horizontal line represents the value of Eq.~(\ref{Eq:ScaleFinal}).
Red solid line: ring without SWBs.
(a) Randomly added SWBs between non-neighboring environment spins.
Green long-dashed line: two SWBs;
orange dotted line: four SWBs;
purple short-dashed line: six SWBs;
blue dotted-dashed line: eight SWBs.
(b) Randomly chosen SWBs between the system and environment spins such that $K=1$.
Green long-dashed line: two SWBs;
orange dotted line: four SWBs;
purple short-dashed line: six SWBs;
blue dotted-dashed line: eight SWBs.
(c) Randomly chosen SWBs between the system and environment spins such that $K=2$.
Green long-dashed line: two SWBs;
orange dotted line: four SWBs;
purple short-dashed line: six SWBs;
blue dotted-dashed line: eight SWBs.
(d) Same as (c) except that each pair of non-neighboring environment spins is connected by a SWB.
Insets: time evolution for short times.
}
\end{figure*}

We investigate the effects of adding small world bonds (SWBs) to the Hamiltonians $H_{SE}$ or/and $H_E$ for
both case~\uppercase\expandafter{\romannumeral1} and case~\uppercase\expandafter{\romannumeral2} (see Fig.~\ref{fig_lat}).
To analyze the addition of SWBs to $H_{SE}$ we distinguish between spin systems with $K<2$ and $K\ge 2$,
where $K$ denotes the maximum number of
subsystem spins that are connected via SWBs with one environment spin.
This distinction is motivated by the distinct decoherence characteristics for systems with $K<2$ and $K\ge 2$ for
case~\uppercase\expandafter{\romannumeral1} (see next subsection).
An example of a spin configuartion with $K=2$ is shown in Fig.~\ref{fig_lat}.
In particular, we are interested in whether systems with SWBs will exhibit the
same scaling, and whether they will decohere from an initial state faster than
either of the cases studied thus far.
The addition of many SWBs changes the graph from a one-dimensional ring to a graph with equal bond lengths that can
only be embedded in high dimensions.
The initial states are always $UDUDY$.
Furthermore, in order not to change too many parameters simultaneously we
start all simulations from the same state ``$Y$'' of the environment. Furthermore, after choosing
the random location (and couplings $\Omega^\alpha$ and $\Delta^\alpha$ for case~\uppercase\expandafter{\romannumeral1})
of the first SWB we preserve this bond when adding additional SWBs.
We will see that case~\uppercase\expandafter{\romannumeral1} and case~\uppercase\expandafter{\romannumeral2} still behave very differently.

\subsection{Case~\uppercase\expandafter{\romannumeral1} and SWBs}

For investigating the universality of the final value of $\sigma(t)$
we add SWBs (random couplings in the interval
$[-0.2,0.2]$) in the Hamiltonian $H_{SE}$ or/and $H_E$
for case~\uppercase\expandafter{\romannumeral1}, and perform
simulations for $N=24$ with $N_S=4$.
From Fig.~\ref{fig_case1_swb}a, we see that adding more and more SWBs to $H_E$ speeds up the decoherence process
and that the final value of $\sigma (t)$ corresponds to the one given by Eq.~(\ref{Eq:ScaleFinal}).
As seen in the inset, adding SWBs to $H_E$ has no noticeable effect on the early time behavior of $\sigma (t)$.

Adding SWBs exclusively to $H_{SE}$ speeds up the decoherence process even further and even at early times clear changes in $\sigma (t)$ can be observed (see Figs.~\ref{fig_case1_swb}b, c).
For spin configurations with $K=1$, $\sigma (t)$ reaches the value given by Eq.~(\ref{Eq:ScaleFinal})
for sufficiently long times, as can be seen from Fig.~\ref{fig_case1_swb}b.
However, for configurations with $K=2$ (see Fig.~\ref{fig_case1_swb}c) or $K>2$ (results not shown) $\sigma (t)$
does not obey the scaling property Eq.~(\ref{Eq:ScaleFinal}).
Restoring this scaling property seems to require an environment that is much more complex than the one-dimensional one as indicated
by Fig.~\ref{fig_case1_swb}d in which we present simulation results for
the case that SWBs between all non-neighboring environment spins have been added.

\subsection{Case~\uppercase\expandafter{\romannumeral2} and SWBs}
For case~\uppercase\expandafter{\romannumeral2}, isotropic SWBs are added to $H_{SE}$ or/and $H_E$.
From Fig.~\ref{fig_case2_swb}, it is clear that even for long times
none of the curves approach the dotted horizontal line, the value of $\sigma(t)$
for an initial state ``$X$".
Adding SWBs exclusively to $H_{E}$ does not have a dramatic effect on $\sigma(t)$ and
has very little effect at early times (see Fig.~\ref{fig_case2_swb}a).

Just as for case~\uppercase\expandafter{\romannumeral1}, it is seen that adding a few SWBs exclusively to $H_{SE}$ for a spin configuration with $K=1$ significantly
decreases the time to approach the steady state, and that the SWBs in $H_{SE}$
also lead to a decrease in $\sigma(t)$ for a fixed time even at early times (see Fig.~\ref{fig_case2_swb}b).
For spin configurations with $K=2$ case~\uppercase\expandafter{\romannumeral1} and case~\uppercase\expandafter{\romannumeral2} seem to have similar
decoherence properties if SWBs are added exclusively to $H_{SE}$, as seen by comparing Fig.~\ref{fig_case1_swb}c and Fig.~\ref{fig_case2_swb}c.
However, connecting in addition each pair of non-neighboring environment spins by isotropic SWBs drives the curves very far away from
the value of $\sigma(t)$ for an initial state ``$X$" (see Fig.~\ref{fig_case2_swb}d).

\subsection{Summary: SWBs}
Adding SWBs to $H_{SE}$ or/and to $H_{E}$ changes the rate of decoherence
as seen by the approach to the asymptotic value for $\sigma(t)$.
In case~\uppercase\expandafter{\romannumeral2}, adding isotropic SWBs to $H_{SE}$ or $H_E$ effectively alters some
spin-spin correlations leading to a decrease in the steady-state value of $\sigma(t)$.
However, this decrease is not sufficient to reach the steady-state value of
$\sigma(t)$ that complies with the prediction Eq.~(\ref{Eq:ScaleFinal}).
Adding isotropic SWBs to $H_{SE}$ and connecting in addition each pair of non-neighboring environment spins by isotropic SWBs drives the curves very far away from
the value of $\sigma(t)$ for an initial state ``$X$", even much further away than the steady-state value for a ring without SWBs.
In contrast to case~\uppercase\expandafter{\romannumeral1} systems with $K<2$ and $K\ge 2$ do not behave significantly different.

Comparing case~\uppercase\expandafter{\romannumeral2} with case~\uppercase\expandafter{\romannumeral1} for $K<2$, we conclude that without introducing the randomness in the
$x,$ $y,$ $z$ components of the spin-spin couplings, the dynamics cannot drive the
system to decoherence if the initial state is different from a state ``$X$".
Increasing the complexity of the environment by adding isotropic SWBs between all non-neighboring environment spins does not help in this respect, even on the contrary.
However, for case~\uppercase\expandafter{\romannumeral1} and configurations with $K\ge 2$, increasing the complexity of the environment
by adding SWBs between all pairs of non-neighboring environment spins allows the dynamics to drive the system to decoherence.

For both case~\uppercase\expandafter{\romannumeral1} and case~\uppercase\expandafter{\romannumeral2}, adding SWBs in $H_{SE}$ and $H_E$ separately speeds up the decoherence in that it evolves more quickly
to a stationary state.
The asymptotic value for $\sigma(t)$ is approached much faster when adding SWBs to $H_{SE}$
instead of $H_E$, and the SWBs in $H_{SE}$ also affect $\sigma(t)$ at early times.
Thus a random SWB coupling to the system via $H_{SE}$ is
the most effective way to decrease the time for decoherence.

\section{Randomness in the environment}
\label{sec3d}
\begin{figure}[t]
\includegraphics[width=8cm]{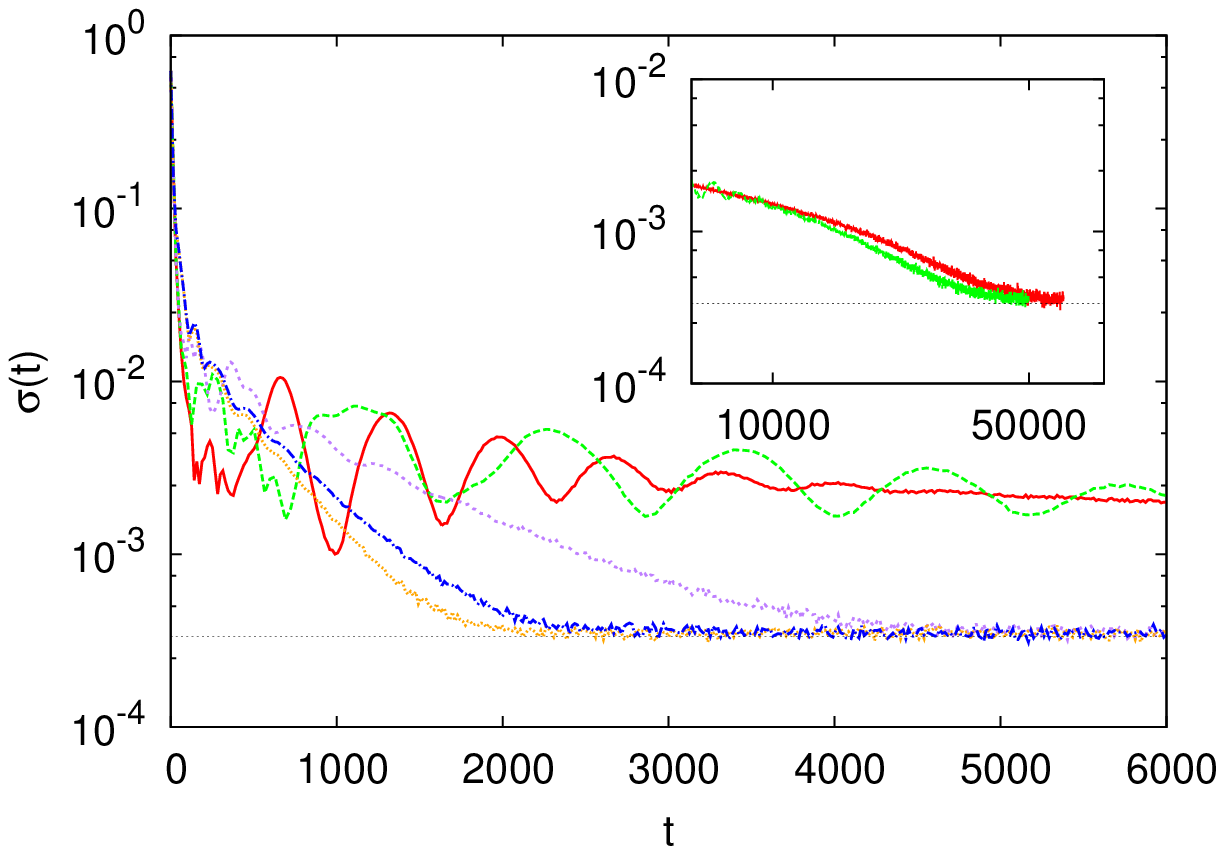}%
\caption{\label{fig_randomness}
(Color online) Simulation results of $\sigma(t)$ obtained by selectively replacing isotropic
spin-spin interactions by random bonds.
The size of the system and whole system are $N_S=4$ and $N=26$, respectively.
The initial state is $UDUDY$.
Red solid line: $1$ random bond; green long-dashed line: $2$ random bonds;
purple dotted line: $4$ random bonds;
orange short-dashed line: $6$ random bonds; blue dotted-dashed line: $8$ random bonds.
Inset: simulation results for one and two random bonds for long times.}
\end{figure}

Section~\ref{sec3a} shows that for the initial state ``$X$" the scaling predicted by
Eq.~(\ref{Eq:ScaleFinal}) is confirmed both for case~\uppercase\expandafter{\romannumeral1} and case~\uppercase\expandafter{\romannumeral2} (see Fig.~\ref{fig_case12}).
However, section~\ref{sec3b} shows that starting from the initial state $UDUDY$
this scaling is approached as $1/\sqrt{D_E}$ for case~\uppercase\expandafter{\romannumeral1} (see Figs.~\ref{fig_case1}
and~\ref{fig_diffsigma}) but not for case~\uppercase\expandafter{\romannumeral2} (see Figs.~\ref{fig_case2}
and~\ref{fig_diffsigma}).
Section~\ref{sec3c} shows that adding SWBs in case~\uppercase\expandafter{\romannumeral2} does not significantly
change the long-time behavior of $\sigma(t)$ approaching the
predicted value of Eq.~(\ref{Eq:ScaleFinal}).
Therefore the natural question to ask is how much randomness is required for $\sigma(t)$ to obey
the scaling relation Eq.~(\ref{Eq:ScaleFinal}).
To answer this question, we start from the isotropic Heisenberg ring (case~\uppercase\expandafter{\romannumeral2})
and replace the interaction strengths of a few randomly chosen
bonds by random $\Omega_{i,j}^\alpha$ (see Eq.~(\ref{HAME})).

Figure~\ref{fig_randomness} presents the simulation results for $\sigma(t)$ by introducing
$1$, $2$, $4$, $6$ and $8$ random bonds in the environment Hamiltonian $H_E$ of Eq.~(\ref{HAME}).
The interaction strengths $\Omega_{i,j}^\alpha$ of these randomly selected bonds are drawn randomly from a uniform distribution in $[-0.2,0.2]$.
Furthermore, the randomly selected bond for the case with $1$ random bond is also a random bond
for the case with $2$ and more randomly chosen bonds, thereby not changing too many
parameters at a time.
Simulations up to time $t=6000$ show that introducing $4$, $6$ and $8$ random bonds
leads the system to relax to the predicted value of $\sigma$ (see Eq.~(\ref{Eq:ScaleFinal})).
For times up to $t=6000$ the effect of one or two random bonds is not apparent.
Therefore for these two cases we performed extremely long runs as shown in the
inset of Fig.~\ref{fig_randomness}.
The inset shows that even one random bond suffices to recover the asymptotic value Eq.~(\ref{Eq:ScaleFinal}).
However the time scale to reach the asymptotic value of $\sigma$ can become extremely long.
We leave the question of how fast the approach to the predicted value of
$\sigma$ is for future study.

\begin{figure}
  \begin{center}
    \begin{tabular}{cc}
      \resizebox{40mm}{!}{\includegraphics{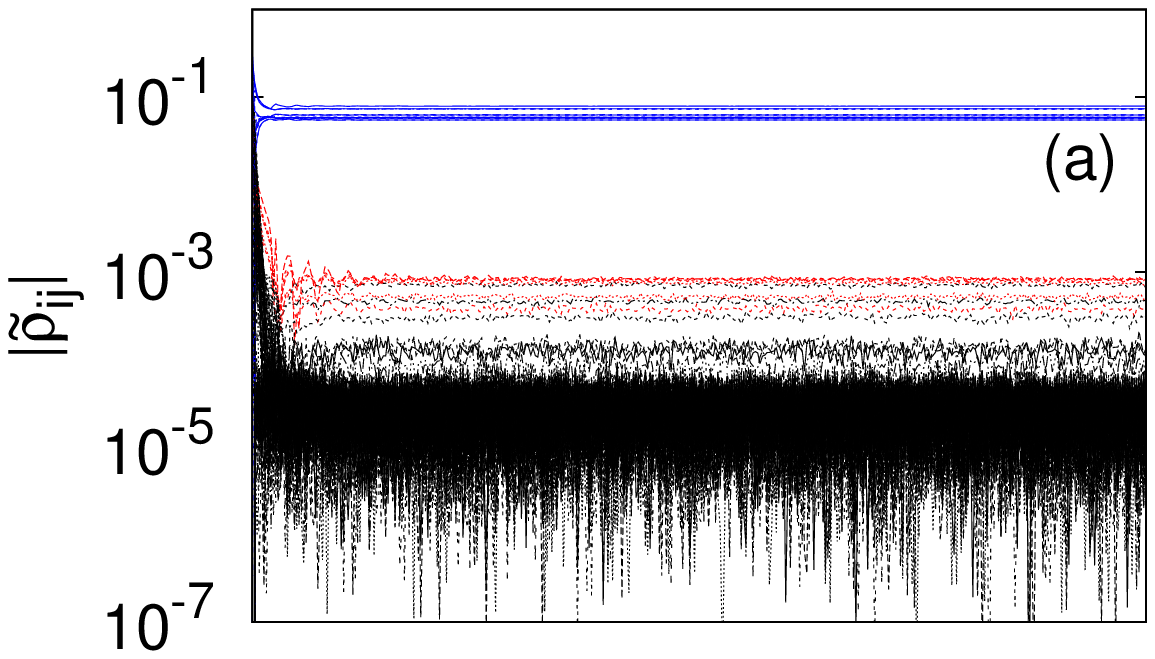}} &
      \resizebox{40mm}{!}{\includegraphics{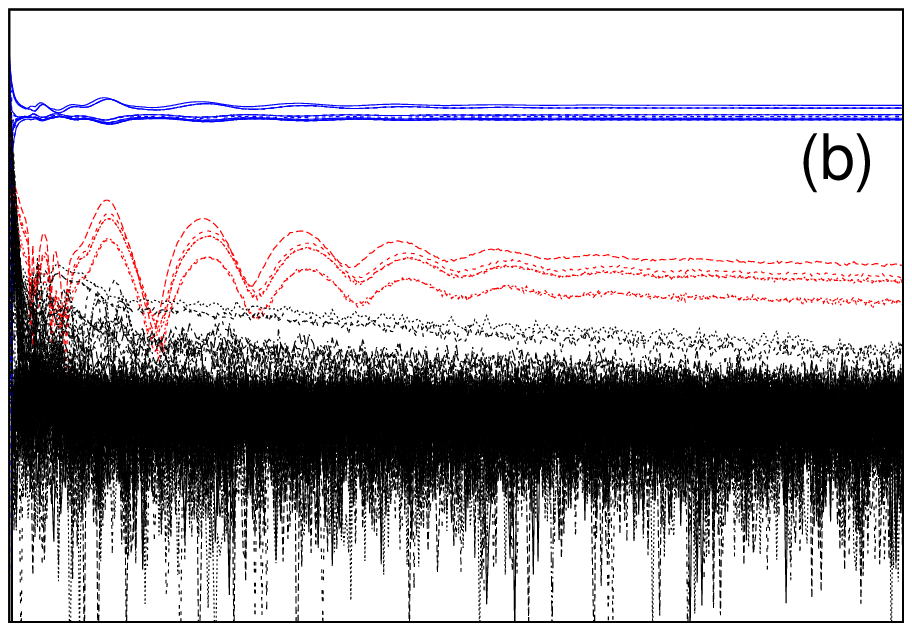}} \\
      \resizebox{40mm}{!}{\includegraphics{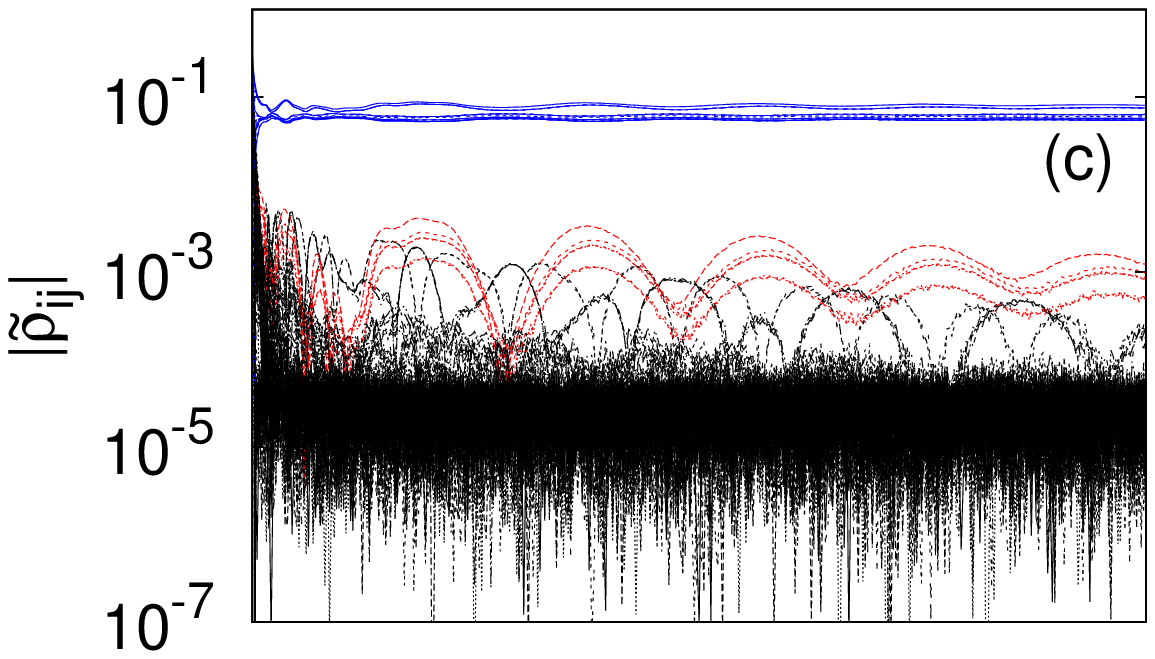}} &
      \resizebox{40mm}{!}{\includegraphics{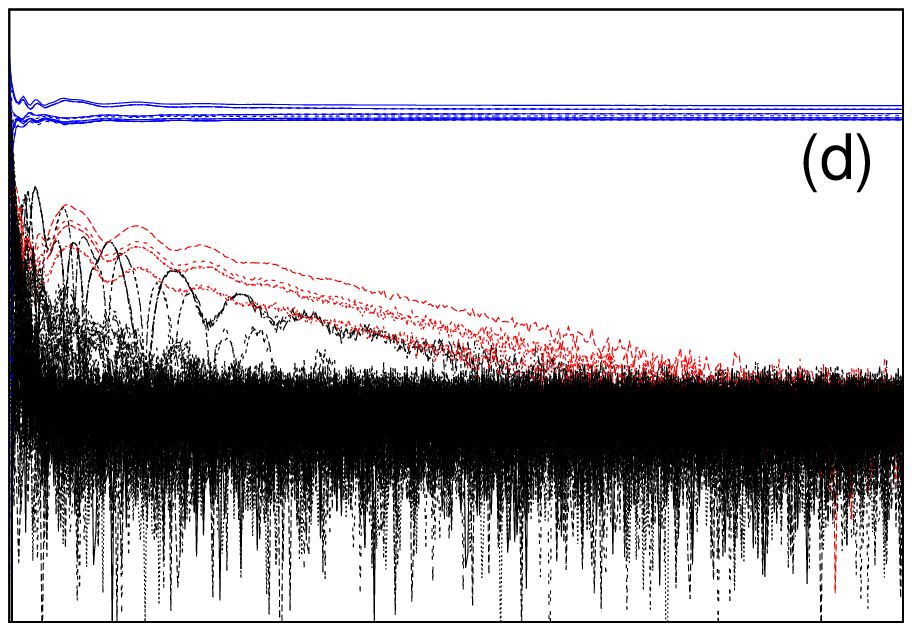}} \\
      \resizebox{40mm}{!}{\includegraphics{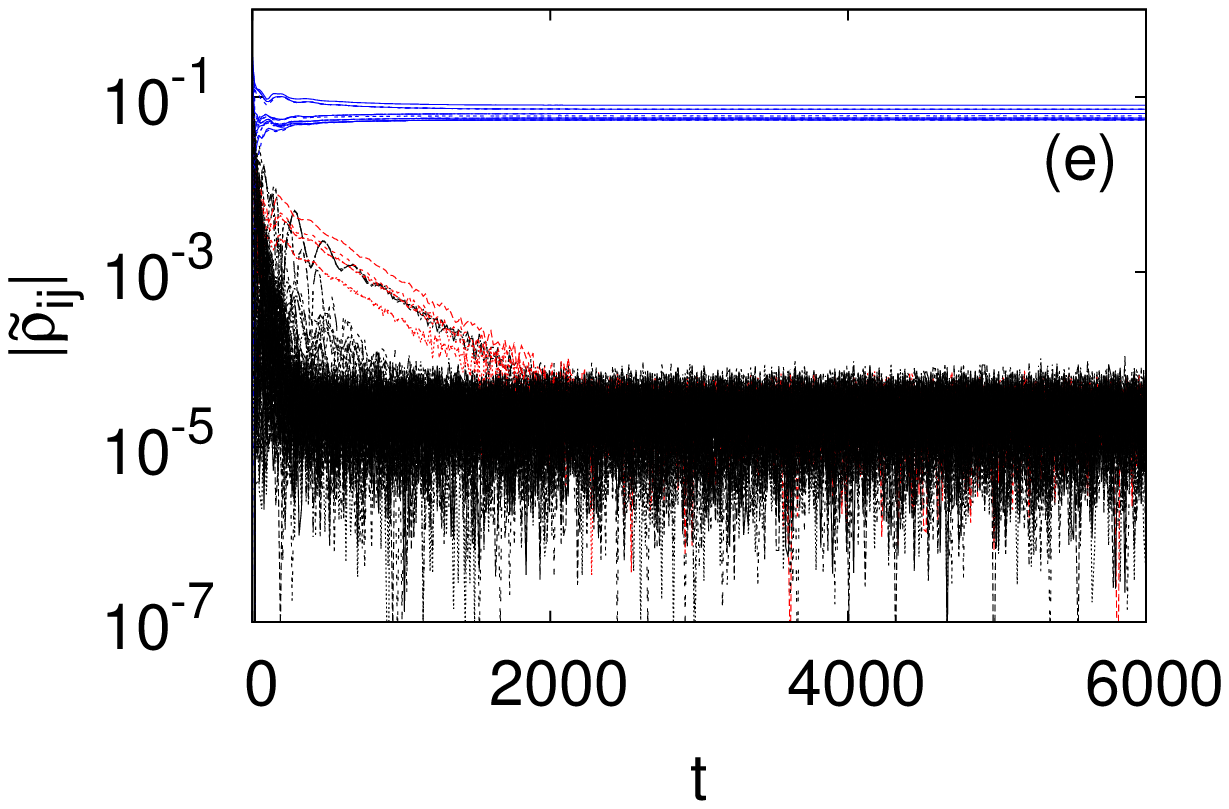}} &
      \resizebox{40mm}{!}{\includegraphics{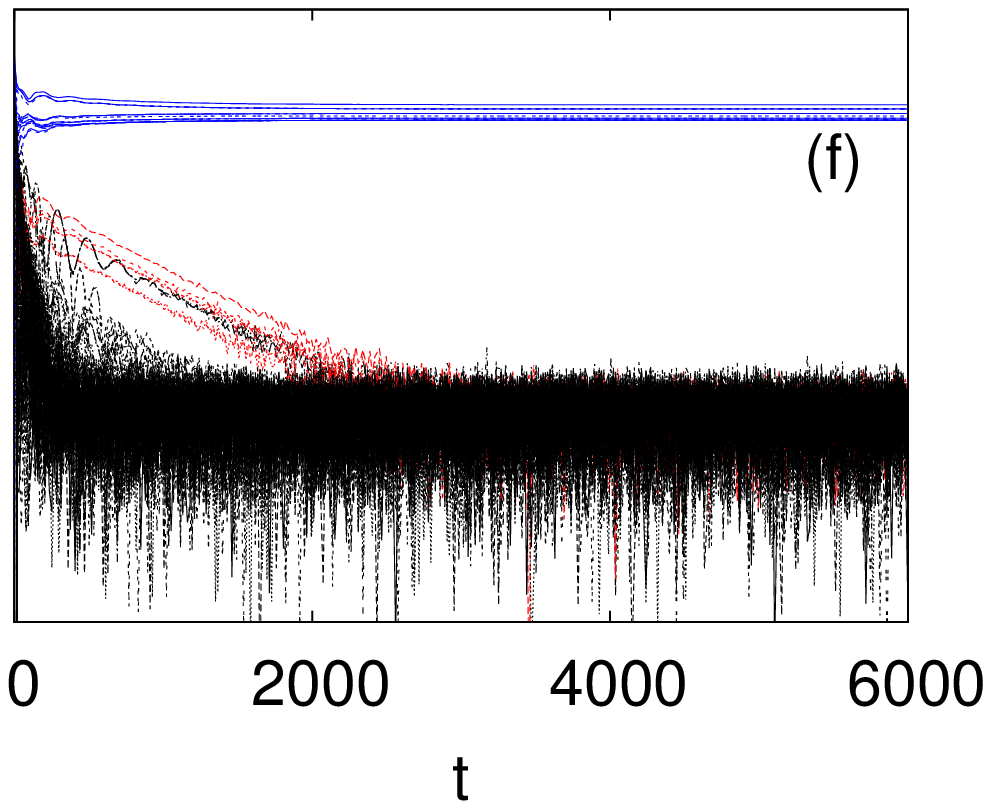}} \\
    \end{tabular}
    \caption{\label{fig_rhoij}
    (Color online) Time evolution of the components $\widetilde\rho_{ij}$ of the
reduced density matrix of the system with $N=26$ and $N_S=4$.
The initial state is $UDUDY$.
    (a) Case~\uppercase\expandafter{\romannumeral2}.
    Starting from case~\uppercase\expandafter{\romannumeral2}, $1$ (b), $2$ (c), $4$ (d), $6$ (e) and $8$ (f) random bonds are introduced in $H_E$.
    Blue lines: diagonal components $|\widetilde\rho_{ii}|$;
    red lines: all $6$ slowly decaying components for $|\widetilde\rho_{ij}|$ for one random bond;
    black lines: all other $114$ off-diagonal components $|\widetilde\rho_{ij}|$.}
  \end{center}
\end{figure}

For understanding the behavior of $\sigma(t)$ in case~\uppercase\expandafter{\romannumeral2} with randomness, we investigate
the individual components of the reduced density matrix $\widetilde\rho$ for the ring system.
We study the addition of one, two, up to eight
randomly replaced bonds in the environment.
Recall that once the position for one random bond is chosen, this is also one
of the random bonds when there are two or more random bonds.
Similarly, the locations of the random positions for a large number of random bonds include the same positions and strengths as for a
smaller number of random bonds.
Furthermore, the same initial state ``$Y$'' of the environment is chosen for all simulations.
We studied the effect of varying the positions of the randomly chosen bonds and
of different initial states ``$Y$'' for the environment for a couple systems and did not find significant changes in our observations.

Figure~\ref{fig_rhoij} presents the results of the time evolution of the absolute value
$|\widetilde\rho_{ij}|$ of the individual components of the reduced
density matrix.
For completeness we show both the diagonal components and the off-diagonal components.
Figure~\ref{fig_rhoij} shows that most of the $120$ off-diagonal components quickly relax to a
small value ($114$ black lines in Fig.~\ref{fig_rhoij} (b)-(e)).
The slowest decaying $|\widetilde\rho_{ij}|$ are plotted in red. There are six such components.
In the steady state all $|\widetilde\rho_{ij}|$ oscillate but have nearly
the same time-averaged value, in agreement with the mean-field-type argument given in
Appendix~\ref{AppB}.
Thus, only a few $|\widetilde\rho_{ij}|$ are responsible for the lack of
scaling of $\sigma$ in case~\uppercase\expandafter{\romannumeral2} when starting from the initial state $UDUDY$,
and also for the long times required to approach the predicted value of Eq.~(\ref{Eq:ScaleFinal})
of $\sigma(t)$ in the case that there are one or two random bonds.

\section{Conclusions and Discussion}

The main theoretical result of the current paper is
Eq.~(\ref{Eq:ScaleFinal}) for the decoherence of a
quantum system $S$ coupled to a quantum environment $E$.
For studying decoherence we examine $\sigma(t)$, which is the
square root of the sum of all the off-diagonal elements of
the reduced density matrix $\widetilde\rho$ for $S$ in the basis that diagonalizes the Hamiltonian $H_S$ of the system $S$.
We find (see also Eq.~(\ref{Eq:ScaleFinal})) that
\begin{equation}
\label{eqsigma12}
\sigma \approx \frac{1}{\sqrt{2 D_E}}\left(1-\frac{1}{2 D_S}\right),
\end{equation}
where the reduced density matrix $\widetilde\rho$ for $S$ is
a $D_S\times D_S$ matrix while the
density matrix of the whole system $S+E$ is a
$D\times D$ matrix with $D=D_S D_E$.
Thus $D_E$ does not have to be very large in order for the
predicted scaling to hold, in particular the scaling requires
$D_E\gg 1\gg D_S^{-1}$.  In addition the scaling requires
that $S+E$ is driven from an initial wave function toward a steady state which is
well described by a  state which we called ``$X$''.

We have performed large-scale real-time simulations of the
time-dependent Schr{\"o}dinger equation for $N_S$ spins in the system
and $N_E$ spins in the environment.  We have simulated spin-$1/2$ systems with
$N=N_S+N_E$ up to $N=34$, all with $N_S=4$.  Starting from a state ``$X$''
for $S+E$ the simulations agree very well with the scaling
prediction Eq.~(\ref{eqsigma12}), as shown in Fig.~\ref{fig_case12}.
In Appendix~\ref{AppC} we demonstrate that in this case not only the off-diagonal
elements of $\widetilde\rho$ obey a scaling relation but also its diagonal elements obey
a scaling relation, although a different one.

Therefore as long as the dynamics drives the initial state to a state ``$Z$" which has similar properties as ``$X$''
the scaling relation Eq.~(\ref{eqsigma12}) should hold.
The next step is to examine under what conditions our test quantum
model is driven to the state ``$Z$'', and study the time scale needed to relax from an
initial state to the state ``$Z$''.  For the one-dimensional quantum spin-$1/2$ ring
we find that homogeneous couplings do not lead to an evolution to the state ``$Z$''
(Fig.~\ref{fig_case2}), and hence the scaling as $1/\sqrt{D_E}$ is not observed.
This conclusion is not modified if some randomly chosen
homogeneous small world bonds are added (Fig.~\ref{fig_case2_swb}).
Also systems with random couplings and random small world bonds between system and
environment spins such that the maximum number of system spins
that interact with one environment spin is two or larger do not evolve to a state ``Z'' (Fig.~\ref{fig_case1_swb}c).
In this case, the environment requires a more complex connectivity than the simple one-dimensional one
in order to observe the scaling as $1/\sqrt{D_E}$ (Fig.~\ref{fig_case1_swb}d).
Therefore, although we find that some randomness in the interaction strengths
in $E$ or between $S$ and $E$ the dynamics is very important to drive the whole system toward the state ``$Z$'',
as seen in Figs.~\ref{fig_case1}, \ref{fig_22spin}, \ref{fig_case1_swb}a,b,
and \ref{fig_randomness} it is not always sufficient.  Moreover it may take a long time to evolve toward
the state ``$Z$'' if there is only a little randomness (Fig.~\ref{fig_randomness})
or if the environment $E$ is large (the $N=34$ results of Fig.~\ref{fig_case1}).
The long time that may be required to approach the state ``$Z$'' is due to
only a few off-diagonal elements of $\widetilde\rho$, as seen in
Fig.~\ref{fig_rhoij}.  We find that the approach to the state ``$Z$'' can be
sped up by adding randomness to $E$ (Figs.~\ref{fig_randomness} and~\ref{fig_rhoij}).

What do our results say about the approach to the quantum canonical ensemble?
The canonical ensemble is given by the diagonal elements of the reduced
density matrix $\widetilde\rho$ if the off-diagonal elements (as measured by $\sigma(t)$)
can be neglected \cite{KUBO85,JIN10b}.
As long as $E$ has a finite Hilbert space $D_E$ our scaling results can be used to
argue that in a strict sense, the system will not be in the canonical state unless $D_E\rightarrow\infty$.
However, if the canonical distribution is to be a good approximation
for some temperatures $T$ up to some chosen maximum energy $E_{\rm hold}>0$, then this
requires that $\exp\left(-E_{\rm hold}/k_B T\right)\gg\sigma$ which gives for
our spin-$1/2$ system
$k_B T\gg 2 E_{\rm hold}/\left[N_E \ln(2)\right]$.
For this argument to hold in the canonical distribution the energies are taken to be
positive values above the ground state energy.
This lack of thermalization at low temperatures for small systems is supported by
simulations in Ref.~\cite{JIN10b}.

What do our results say about trying to prolong the time to decoherence
in order to build practical quantum encryption or quantum computational devices?
The important thing is to ensure that the system is not driven toward the state
``$Z$'', or at least that it takes a very long time to approach
the state ``$Z$''. This can be achieved by changing the
Hamiltonian of the system, $H=H_S+H_E+H_{SE}$, such that it has very small randomness
particularly in the coupling between the system and the environment, $H_{SE}$.
Alternatively extrapolating from Fig.~\ref{fig_rhoij} if one can devise an
experimental procedure, for example a time-dependent procedure, to keep
even a few of the off-diagonal elements of $\widetilde\rho$ large then the
scaling prediction Eq.~(\ref{eqsigma12}) for the decoherence can be avoided, at
least for reasonable timescales.

The scaling of Eq.~(\ref{eqsigma12}) can be contrasted with the predicted scaling
of the Hilbert space variant of a whole system which should be
proportional to $(D+1)^{-1}$ for the expectation value of a local operator~\cite{BART09}.
The results of the current research are also relevant
for methodologies for measuring finite-temperature dynamical correlations~\cite{LONG03}
without performing the complete TDSE evolution of the
whole system.

We leave as future work the coupling between a system $S$ composed of spin-$1/2$ objects
(qubits) and an environment $E$ composed of harmonic oscillators.
In particular, we have recently been able to build on exact calculations
of a single spin coupled to specific types of spin environment \cite{RAO08}
to devise an algorithm that does not have computer memory constraints
limited by the size of $D_E$ \cite{NOVO12a,NOVO12b}.  We are working to extend this
algorithm to other types of environment and for more than one spin in the system $S$.

\appendix
\section{Scaling without an environment}
\label{AppA}
For comparison of the scaling of $\sigma(t)$ for the cases with and without an environment, we derive the scaling for
the case of no environment.  In the energy basis $\left|i\right\rangle$
of the (system, which is now the whole system)
Hamiltonian $H$, the density matrix has elements
\begin{equation}
\rho_{ij}(t) = c_i(t) c_j^\dagger(t)
\>.
\end{equation}
We use from Ref.~\cite{HAMS00} the equations (A.12) and (A.23).
The expectation value is
\begin{widetext}
\begin{eqnarray}
\label{eq_scaling_NoBath}
E\left(2\sigma^2\right)
&=&E\left(\sum_{i=1}^{D_S} \sum_{j\neq i}^{D_S} \left|c_i(t) c_j(t) \right|^2 \right)
=\sum_{i=1}^{D_S} \sum_{j\neq i}^{D_S} E\left(\left|c_i(t) c_j(t) \right|^2 \right) \cr
&=& D_S\left(D_S-1\right) E\left(\left|c_i(t)\right|^2 \left| c_j(t) \right|^2 \right)
= 1-\frac{2}{D_S+1} \>=\> \frac{D_S-1}{D_S+1}
\> .
\end{eqnarray}
The final scaling result for the quantity $\sigma$ that we measure is
\begin{equation}
\label{Eq:ScaleFinal_NoBath}
\sigma \approx
\frac{1}{\sqrt{2}}
\sqrt{E\left(2\sigma^2\right)} =
\frac{1}{\sqrt{2}}\sqrt{\frac{D_S-1}{D_S+1}}
=
\frac{1}{\sqrt{2}} -\frac{1}{\sqrt{2} D_S} +\frac{1}{2\sqrt{2}D_S^2}
-\frac{1}{2\sqrt{2} D_S^3} +\frac{3}{8\sqrt{2}D_S^4} +\cdots
\>.
\end{equation}
\end{widetext}
Therefore without an environment, $\sigma$ approaches a constant as the size of the
system (which is the whole system) grows.
This also means that for the state ``$X$", if all off-diagonal elements are
the same they will have a size of
$\left|\rho_{ij}(t)\right|^2=1/D_S\left(D_S-1\right)\sim 1/D_S^2$
while if all the diagonal elements are equal (corresponding to infinite temperature)
$\left|\rho_{ii}(t)\right|^2=1/D_S$ since ${\rm Tr}~\rho(t)=1$.
We have performed simulations (results not shown)
to ensure that for the case without an environment $\sigma$
obeys the scaling relation of Eq.~(\ref{Eq:ScaleFinal_NoBath}) and it does.

\section{Mean-field-like reduced density matrix}
\label{AppB}

We make a connection between $\sigma$ and the quantum purity
${\cal P}={\rm Tr}\left(\left(\hat\rho\right)^2\right)$.
We assume a `mean-field-type' structure for the reduced density matrix, namely we assume that
all off-diagonal elements have the same size, $\epsilon$.
In our simulations we find that in the energy basis the imaginary part of
the off-diagonal
elements are very small, which validates our hypothesis.  However, the
signs of the real part of the off-diagonal elements are not the same, which brings
into question our `mean-field-like' assumption.
Nevertheless, we make the assumption that
\begin{equation}
\epsilon=\sqrt{\frac{2\sigma^2}{D_S\left(D_S-1\right)}}
\>.
\end{equation}
We introduce the matrix
${\bf J}$ with all its elements having the value $1$, the matrix ${\bf D}$ which is the diagonal matrix
composed of the diagonal elements of ${\hat\rho}$, and the identity matrix ${\bf I}$.
Note that ${\bf J}^2 = D_S {\bf J}$.
The `mean-field-type' assumption then reads
\begin{equation}
{\hat\rho} = {\bf D} + \epsilon{\bf J} - \epsilon{\bf I},
\end{equation}
which as seen from the graphs in Fig.~\ref{fig_rhoij}
should be a reasonable assumption in the steady state regime.
We will use the relationships
\begin{eqnarray}
{\rm Tr}\left({\bf D}\right) & = & 1, \cr
{\rm Tr}\left({\bf D}^2\right) & \le & 1, \cr
{\rm Tr}\left({\bf I}\right) & = & {\rm Tr}\left({\bf J}\right) = D_S, \cr
{\rm Tr}\left({\bf D}{\bf J}\right) & = & {\rm Tr}\left({\bf J}{\bf D}\right) = 1, \cr
{\rm Tr}\left({\bf J}^2\right) & = & D_S^2,
\end{eqnarray}
with the first relationship being a consequence of the trace of a density matrix being equal to unity.
Then one has that
\begin{eqnarray}
{\cal P} & = & {\rm Tr}\left({\hat\rho}^2\right) \cr
& = & {\rm Tr}\left(\left({\bf D} + \epsilon{\bf J} - \epsilon{\bf I}\right)^2\right) \cr
& = & {\rm Tr}\left({\bf D}^2 -2\epsilon{\bf D}+\epsilon^2{\bf I}+\epsilon{\bf D}{\bf J}
      +\epsilon{\bf J}{\bf D}-2\epsilon^2{\bf J} + \epsilon^2{\bf J}^2\right) \cr
& = & {\rm Tr}\left({\bf D}^2\right) + 2\sigma^2 \cr
& = & {\rm Tr}\left({\bf D}^2\right) + \frac{1-\frac{1}{D_S}}{D_E+\frac{1}{D_S}} \cr
& = & {\rm Tr}\left({\bf D}^2\right) \cr
&& + \frac{1}{D_E}\left(1-\frac{1}{D_S}-\frac{1}{D_E D_S}
      +\frac{1}{D_E D_S^2}+\cdots\right)
\>.
\end{eqnarray}
In the canonical ensemble the diagonal elements of the reduced density matrix
are related to the
terms in the canonical partition function, in particular
${\hat\rho}_{ii} = e^{-\beta E_i}/Z$~\cite{YUAN09,JIN10b}.
Therefore we have a connection between the quantum purity
${\cal P}$ and how close the system is to a canonical ensemble.
In the steady state this difference is of the order of $1/D_E$.

With the same `mean-field-like' assumption for ${\hat\rho}$ in the steady
state one can look at corrections to the von~Neumann entropy of the system,
${\cal S}=-{\rm Tr}\left({\hat\rho}{\rm ln}{\hat\rho}\right)$.
However, we do not find the final result too enlightening.

\section{Diagonal elements of the reduced density matrix}
\label{AppC}
In the main text, we investigated the scaling property of the off-diagonal elements
of the reduced density matrix of a system coupled to an environment.
For being complete in the contents, we present some numerical and analytical results concerning
the diagonal elements.

In general, based on the fact that the system decoheres, i.e.\ the off-diagonal elements of the reduced density matrix approach zero,
we expect that the diagonal elements take (approach to) the form of the canonical distribution
$\exp(-\beta E_{i})$ where $\beta=1/k_BT$ with $T$ denoting the temperature and
$k_B$ Boltzmann's constant, which is taken to be one in this paper,
and where $E_{i}$'s denote the eigenvalues of $H_S$~\cite{YUAN09,JIN10b}.
The difference between the diagonal elements $\widetilde\rho_{ii}\left( t\right)$
and the canonical distribution is conveniently characterized by
\begin{equation}
\delta(t)=\sqrt{\sum_{i=1}^{D_S}\left( \widetilde\rho_{ii}(t) -
\left.{e^{-b(t) E_{i}}}\right/{\sum_{i=1}^{D_S} e^{-b \left( t\right) E_{i}}}\right) ^{2}}
,
\label{eqdelta}
\end{equation}%
with a fitting inverse temperature
\begin{equation}
b(t)=\frac{\sum_{i<j,E_{i}\neq E_{j}}
[\ln \widetilde\rho_{ii}(t) -\ln \widetilde\rho _{jj}(t)]/({E_{j}-E_{i}})}{\sum_{i<j,E_{i}\neq E_{j}}1}.
\label{eqbt}
\end{equation}%
If the system relaxes to its canonical distribution
both $\delta(t)$ and $\sigma (t)$ are expected to vanish,
$b(t)$ converging to the effective inverse temperature $b$.

\begin{figure}[t]
\includegraphics[width=8cm]{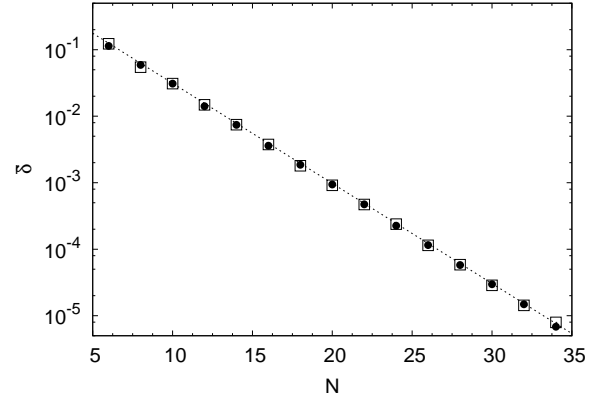}
\caption{\label{fig_delta12}Simulation results for the time-averaged value $\bar{\delta}$ of $\delta(t)$
(see Eq.~(\ref{eqdelta})) for case~\uppercase\expandafter{\romannumeral1} (bullets) and
case~\uppercase\expandafter{\romannumeral2} (squares) for different sizes $N$ of the whole system.
The initial state of the whole system is ``$X$" (see text).
The dotted line is $1/\sqrt{2^N}$.
}
\end{figure}

The numerical simulations of which we present the results correspond to those
used to make Fig.~\ref{fig_case12}.
The initial state for those simulations is ``$X$".
We analyze the diagonal elements, instead of the off-diagonal elements, of the reduced density matrix and calculate the quantity $\delta(t)$.
In Fig.~\ref{fig_delta12}, we present the time-averaged value $\overline{\delta}$ of $\delta(t)$ for each system size.
It is interesting to see that the quantity $\overline{\delta}$ also has a kind of scaling property.
As the whole system size $N$ increases, $\overline{\delta}$ decreases as $1/\sqrt{D}$, where $D=2^N$.

In fact the fitting inverse temperature $b(t)$ is very close to zero for reasonablely large $N_E$ (data not shown).
The canonical distribution of $S$ at $b=0$ is
represented by a diagonal density matrix with elements $1/D_S$, where $D_S=2^{N_S}$.
Then, we are able to derive the scaling property for $\delta$ as we did to obtain Eq.~(\ref{eq_scaling}).
The expectation value of $\delta$ is given by
\begin{widetext}
\begin{eqnarray}
\label{eq_delta}
E\left(\delta^2\right)
&=&E\left(\sum_{i=1}^{D_S} \left |\sum_{p=1}^{D_E} C_{i,p}^* C_{i,p}-\frac{1}{D_S} \right|^2\right)
 = \sum_{i=1}^{D_S} \> \sum_{p=1,p'=1}^{D_E} E\left(
    \left |C_{i,p}\right |^2 \left | C_{i,p'} \right |^2\right)-\frac{1}{D_S} \cr
&=&\sum_{i=1}^{D_S} \> \sum_{p=1,p'=1}^{D_E} \left( \left(1-\delta_{p,p'}\right)
 E\left(\left |C_{i,p}\right |^2 \left | C_{i,p'} \right |^2\right)+\delta_{p,p'}
 E\left(\left |C_{i,p}\right |^4\right) \right) -\frac{1}{D_S}\cr
&=&\sum_{i=1}^{D_S} \> \sum_{p=1,p'=1}^{D_E} \left( \left(1-\delta_{p,p'}\right)
 \frac{1}{D(D+1)}+\delta_{p,p'} \frac{2}{D(D+1)} \right) -\frac{1}{D_S}\cr
 &=& \frac{D_E+1}{D+1}-\frac{1}{D_S}=\frac{D_S-1}{D_S}\frac{1}{D+1}
\> .
\end{eqnarray}
\end{widetext}
From Eq.~(\ref{eq_delta}), we have $\delta \approx 1/\sqrt{D}$ for $D_S> 1$ and $D_E\gg 1$.
Therefore, if the size of the environment goes to infinity with the final state being the state ``$X$", the diagonal elements of the reduced density matrix of the system approach $1/D_S$.

\section*{Acknowledgements}
This work is supported in part by NCF, The Netherlands (HDR),
the Mitsubishi Foundation (SM), and the US National Science Foundation under Grant No. DMR-1206233 (MAN).
MAN acknowledges support from the J\"ulich Supercomputing Centre.
Part of the calculations has been performed on JUGENE and JUQUEEN at JSC under VSR project 4331.

\bibliography{../scaleH,../../../epr11}

\end{document}